\documentclass[aps,twocolumn,superscriptaddress,nofootinbib,pre]{revtex4-2}
\usepackage[pdftex]{graphicx}
\usepackage{latexsym,amsmath,verbatim}
\usepackage{xcolor}
\usepackage[hidelinks,unicode=true]{hyperref}
\hypersetup{colorlinks=true,linkcolor=blue,urlcolor=blue,citecolor=blue,pdfhighlight=/N
}
\usepackage{lipsum}
\usepackage[english]{babel}
\usepackage{microtype}
\usepackage[normalem]{ulem}

\newcommand{\av}[1]{\left\langle {#1} \right\rangle}
\newcommand{\kmax}{k_\mathrm{max}}
\newcommand{\kmin}{k_\mathrm{min}}
\newcommand{\be}{\begin{equation}}
\newcommand{\ee}{\end{equation}}

\graphicspath{{./figures/}}

\begin{document}
\title{Influence of individual nodes \\ for continuous-time Susceptible-Infected-Susceptible
  dynamics \\ on synthetic and real-world networks}

\author{Alfredo De Bellis}
\affiliation{Dipartimento di Fisica,
Sapienza Universit\`a di Roma, P. le A. Moro 2, I-00185 Roma, Italy}

\author{Romualdo Pastor-Satorras}
\affiliation{Departament de F\'{\i}sica, Universitat Polit\`ecnica de
Catalunya, Campus Nord B4, 08034 Barcelona, Spain}

\author{Claudio Castellano}
\affiliation{Istituto dei Sistemi Complessi (ISC-CNR), Via dei Taurini 19,
I-00185 Roma, Italy}


\begin{abstract}
  In the study of epidemic dynamics a fundamental question is whether
  a pathogen initially affecting only one individual will give rise to
  a limited outbreak or to a widespread pandemic. The answer to this
  question crucially depends not only on the parameters describing the
  infection and recovery processes but also on where, in the network
  of interactions, the infection starts from.  We study the dependence
  on the location of the initial seed for the
  Susceptible-Infected-Susceptible epidemic dynamics in continuous
  time on networks. We first derive analytical predictions for the
  dependence on the initial node of three indicators of spreading
  influence (probability to originate an infinite outbreak, average
  duration and size of finite outbreaks) and compare them with
  numerical simulations on random uncorrelated networks, finding a
  very good agreement.  We then show that the same theoretical
  approach works fairly well also on a set of real-world topologies of
  diverse nature. We conclude by briefly investigating which
  topological network features determine deviations from the
  theoretical predictions.
\end{abstract}

\maketitle

\section{Introduction}

As the present pandemic keeps reminding us, disease spreading plays in our
world a role that can hardly be overestimated~\cite{christakis2020apollo}.
To understand how an infectious pathogen in a single individual develops
into a widespread epidemic it is crucial to investigate the relationship
between the topology of the interaction pattern and the collective
properties of the contagion~\cite{PastorSatorras2015, Kiss2017}.  Among the
nontrivial questions that spreading phenomena in complex networks raise, the
effect of the exact location where the contagion is seeded is one of the
most interesting.  Is the course of an epidemic strongly depending on where
the pathogen first appears? Is it possible to calculate this dependence,
thus identifying a priori influential spreaders to be monitored?  Are purely
topological quantities (centralities) able to provide information on the
size or duration of an outbreak started in a given node?  After the seminal
paper by Kitsak et al.~\cite{Kitsak2010} these questions have attracted a
lot of interest.  The natural framework for investigating these issues is
the stochastic Susceptible-Infected-Removed (SIR) model, where susceptible
nodes can be infected by neighboring nodes in a network and then
spontaneously recover, acquiring permanent immunity.  Spreading influence is
naturally measured by considering the average final size $S_n$ of an
outbreak originated by a seed placed in node $n$.  Such a quantity is finite
(on any finite network) for any value of the parameters describing the
dynamics.  Initially, the focus has been on the identification of
topological centralities, such as degree, K-core or eigenvalue centrality,
sufficiently correlated with outbreak size, in the sense that ranking nodes
based on the centrality provides the correct ranking also for what concerns
$S_n$.  Many methods and associated quantities have been considered to
perform this goal~\cite{Lu2016}.  The mapping of SIR static properties to
bond percolation has allowed to recognize Non-Backtracking centrality as the
solution of the problem for random networks at
criticality~\cite{Radicchi2016} and to calculate, via message-passing
methods, the spreading influence throughout the whole
phase-diagram~\cite{Min2018}.

Concerning instead epidemic processes without permanent immunity (such as
the Susceptible-Infected-Susceptible, SIS) activity has been much more
limited~\cite{Liu2017,Holme2018}, because in this case the very definition
of spreading influence is not trivial, since above the epidemic threshold a
fraction of all outbreaks lasts indefinitely in time.  Very recently, we
have tackled this issue and studied the problem of influential spreaders for
SIS dynamics in discrete time~\cite{PouxMedard2020}.  In that paper, we have
considered three different quantities that measure spreading influence and
applied an existing theoretical framework~\cite{Larremore2012} (conceptually
equivalent to the quenched mean-field approach~\cite{PastorSatorras2015}) to
calculate them analytically. Numerical simulations, performed on random
uncorrelated networks, have confirmed the validity of the approach,
revealing a satisfactory overall agreement with the theoretical predictions.

While the work in Ref.~\cite{PouxMedard2020} constitutes a first systematic
approach to the study of spreading influence for SIS dynamics in networks,
some questions remain open.  The first has to do with the extension of the
results to SIS in continuous time, which is, in many respects, a more
realistic model for infectious disease.  The framework of
Ref.~\cite{Larremore2012} is easily applicable to a discrete time dynamics
where, in particular, the duration of the infected state is set
deterministically to 1 for all individuals.  Because of that, at the same
time a node gets infected, the node that transmitted the infection to it
(infector) necessarily becomes susceptible again. This is very different
from what happens when the duration of the infected state is a random
variate (as in continuous-time SIS): in such a case the infector remains
infected for some time after the infection event, so that the effective
degree of the newly infected individual for further spreading the disease is
temporarily reduce by one.  These dynamical correlations among the state of
neighbors make the application of the approach of Larremore et al.\ to SIS in
continuous time nontrivial and in principle less accurate.

Another relevant question left completely open is the validity of the
approach beyond random uncorrelated networks. Real-world topologies are in
general very different from the synthetic networks considered in
Ref.~\cite{PouxMedard2020}, as they may include arbitrary degree
distributions and correlations, local clustering, mesoscopic structures
(communities, core-periphery) and so on.  Is the theoretical approach able
to accurately describe spreading influence also for such interaction
patterns? How do different topological properties affect the predictive
power of the theory?

In this paper we tackle these major questions, providing an exhaustive
response to the first and, by investigating the SIS model on a set of
real-world structures, highlighting the different role that some topological
properties play in determining the validity of the approach on generic
networks.

\section{Theoretical approach}
\label{sec:mean-field-theory}

We consider the standard SIS dynamics in continuous time on an undirected
unweighted network.  Each node $i$ can be either susceptible or infected.
Infection and healing dynamics are ruled by two Poisson processes. At rate
(probability per unit time) $\mu$  each infected node recovers
spontaneously, becoming susceptible again.  An infected node may transmit
the contagion to each of its susceptible neighbors.  This occurs
independently at rate $\beta$ for each one of the susceptible neighbors.  We
consider an initial condition with a single node $n$ infected, while all the
others are in the susceptible state and we are interested in calculating the
dependence on the seed node $n$ of the probability $b_n$ to observe a finite
outbreak, of the average duration $T_n$ and of the average size $S_n$ of
finite outbreaks~\cite{PouxMedard2020}.

\subsection{Quenched Mean-Field theory}
\label{sec:quenched-field-theory}

Inspired by the approach developed by Larremore \emph{et al.} in
Ref.~\cite{Larremore2012}, we write an equation for the probability $c_n(t)$
that an outbreak starting from a seed node  $n$ has a duration smaller than
or equal to $t$.  We compute it by means of a one-step calculation,
considering the system with only node $n$ infected at time $t=0$ and making
the assumption of a \textit{locally tree-like} network, so that we can
consider outbreaks propagating to the different neighbors of $n$ as
independent. To do so, we consider the probability $c_n(t + dt)$ that a seed
node $n$ generates an outbreak of duration smaller than or equal to
$t +dt$, with  $dt$ an infinitesimal time interval.  Due to the Poisson nature
of the infection and healing processes, in the interval $dt$ the node $n$
can heal with probability $\mu dt$ and can infect each of its $k_n$ nearest
neighbors with probability $\beta dt$. Otherwise, with probability $1-\mu
dt- k_n \beta dt$, nothing happens.  If a recovery event takes place, then
the outbreak stops and consequently its duration is necessarily smaller
than any $t'>dt$.  If the node infects one of its nearest neighbors, the
newly infected node $m$, together with the seed node $n$, still infected,
can give rise to another outbreak, starting at time $t=dt$ from the pair
$(n,m)$ of infected neighbors. For the global outbreak to be shorter than
$t+dt$, this induced outbreak must have a duration shorter than $t$.  If
nothing happens, the node $n$ will still be infected at time $dt$, so we
must impose that the subsequent outbreak it generates has a duration not
larger than $t$.  Denoting by $c_{nm}(t)$ the probability that the adjacent
pair of infected nodes $(n,m)$ generate an outbreak of duration not larger
than $t$, we can write
\begin{eqnarray}
  c_n(t+dt) &=& \mu dt + \bigl[ 1-(\mu+ k_n \beta)dt \bigr]c_n(t) \nonumber
         \\ &+& \sum\limits_m a_{nm}\beta dt c_{mn}(t), 
\end{eqnarray}
where $a_{nm}$ is the adjacency matrix of the network, and we assume that
the duration probabilities $c_n(t)$ and $c_{nm}(t)$ are time translation
invariant.
Rearranging the terms and taking the limit $dt \rightarrow 0$ we arrive at
the equation
\begin{equation}\label{eq_dc}
  \frac{dc_n(t)}{dt}=\mu - \bigl[ \mu + k_n\beta \bigr]c_n(t) + \beta
  \sum\limits_m a_{nm}c_{mn}(t).
\end{equation}

We note that this equation is more complicated than the corresponding
equation for $c_n$ in the case of discrete time dynamics with unit recovery
time treated in Ref.~\cite{PouxMedard2020}.
In that case a node necessarily heals immediately after infecting a
neighbor and for that reason the equation for $c_n$ depends only on $c_m$.
Here instead Eq.~\eqref{eq_dc} for $c_n$ depends on $c_{nm}$.
This is a consequence of the fact that $n$ can be still infected at time $t+dt$,
so dynamical correlations unavoidably arise: $m$ can infect, in the successive
dynamical event, only $k_m-1$ neighbors, instead of $k_m$.
Nevertheless we neglect in the following
these correlations assuming the factorized form $c_{nm}(t)\approx
c_n(t)c_m(t)$, so that the equation for $c_n(t)$ finally reads 
\begin{equation}
  \label{eq_c}
  \frac{dc_n(t)}{dt}=\mu - \bigl[ \mu + k_n\beta \bigr]c_n(t) + \beta c_n(t)
  \sum\limits_m a_{nm}c_m(t),
\end{equation}
to be integrated with the initial condition $c_n(0)=0$. As we will see in
Sec.~\ref{seq:numeric_synthetic}, numerical evidence backs up the
factorization assumption for the probability $c_{nm}(t)$.

\subsection{Probability that an outbreak is finite}
\label{p_b}

The probability of observing a finite outbreak starting from the single seed
node $n$ is given by $b_n=\lim_{t \rightarrow \infty}c_n(t)$.  Imposing the
stationarity condition $\dot{c}_n=0$ in Eq.~(\ref{eq_c}), $b_n$ can be
obtained by solving iteratively the self-consistent equation:
\begin{equation}\label{eq_b_qmf}
  b_n=\frac{\mu + \beta b_n \sum_m a_{nm} b_m}{\mu + k_n \beta} .
\end{equation}
In Appendix~\ref{cap_fixed_points} we show that all the $b_n$ are equal to
$1$ when $\lambda \equiv (\beta/\mu) \Lambda_M \leq 1$ (where $\Lambda_M$ is
the largest eigenvalue of the adjacency matrix) while the fixed point
$b_n=1$ loses its stability, and thus another solution $b_n<1$ appears, when
$\lambda>1$.

\subsection{Average duration of a finite outbreak}

To compute the average duration $T_n$ of a finite outbreak seeded in node
$n$, we consider the outbreak duration distribution $P_n(t)$, that is given
by
\begin{equation}
  P_n(t)= \frac{1}{b_n} \frac{dc_n(t)}{dt} ,
\end{equation}
where the prefactor guarantees normalization, since we are only considering
finite outbreaks~\footnote{We notice that this prefactor was incorrectly
omitted in Ref.~\cite{PouxMedard2020}}.

Let us define $c_n(t)=b_n-f_n(t)$, with $f_n(0)=b_n$ and
$\lim_{t \rightarrow \infty} f_n(t)=0$.
We can thus write
\begin{eqnarray}
  T_n 
  &=& \int_0^{\infty} t' P_n(t')dt' = - \frac{1}{b_n} \int_0^{\infty} t'
  \frac{df_n(t')}{dt'}dt' \nonumber \\
  &=& \frac{1}{b_n} \int_0^{\infty} f_n(t') dt' .
\label{eq_T_qmf}
\end{eqnarray}
Rewriting Eq.~\eqref{eq_dc} in terms of $f_n(t)$ we obtain
\begin{eqnarray}
    -\frac{df_n(t)}{dt} 
  &=& \mu \frac{f_n(t)}{b_n} - \beta b_n \sum_m a_{nm} f_m(t) \nonumber \\
  &+& \beta f_n(t) \sum_m a_{nm} f_m(t),\label{eq_df}
\end{eqnarray}
where we have used the steady state condition Eq.~\eqref{eq_b_qmf}.  By
solving numerically this equation and plugging $f_n(t)$ back into
Eq.~\eqref{eq_T_qmf}, the average duration of a finite outbreak seeded in
$n$ can be evaluated.

In Ref.~\cite{PouxMedard2020}, a similar equation for the average outbreak
size was simplified by performing a linearization, based on the assumption
that $f_n(t)$ decays to zero very quickly, in such a way that we can
disregard the last term in the right hand side of Eq.~\eqref{eq_df}.
While in the discrete time case such approximation
works~\cite{PouxMedard2020}, in continuous time it fails. The failure can be
traced back to the linearized equation being unphysical at intermediate
times for sufficiently large values of the degree $k_n$ (see
Appendix~\ref{unphysical}).

\subsection{Average size of a finite outbreak}

For predicting the average outbreak size, we have to slightly modify
the generating function approach developed in Ref.~\cite{Larremore2012} in
order to take into account once again the dynamical correlations among
nearest neighbors.  Let us define $y_n$ as the final size of an outbreak
starting from node $n$.  The following relation is satisfied by $y_n$:
\begin{equation}\label{yn}
y_n=\prod\limits_m (1-z_{nm})+\sum\limits_m z_{nm}y_{nm},
\end{equation} 
where $z_{nm}$ is a random variable with value $1$ if $n$ has infected node
$m$, and $0$ otherwise, and $y_{nm}$ is the size of the outbreak generated
by the infected pair of adjacent nodes $(n,m)$~\footnote{Note that in the
  discrete-time version, the first term in the right hand side of
  Eq.~\eqref{yn} is replaced by $1$, since in that case, after the first
  event, the seed node necessarily recovers; in continuous time instead the
  size $y_n$ is equal to $1$ if and only if $z_{nm}=0$ for all $m$, i.e., if
the seed node $n$ heals before infecting any of its nearest neighbors.}.  We
restrict our attention only to finite outbreaks and we introduce the moment
generating functions relative to their distribution, as
\begin{equation}\label{eq_phi}
  \phi_n(s) \equiv E \bigl[ e^{-s y_n} | y_n < \infty \bigr],
\end{equation}
and 
\begin{equation}
  \phi_{nm}(s) \equiv E \bigl[ e^{-s y_{nm}} | y_{nm} < \infty \bigr],
\end{equation}
where the expectation value is calculated over the realizations
of the random pairs $(z_{nm}, y_{nm})$.

We rewrite the condition $y_n < \infty$ in terms of the events that $n$
could generate. An outbreak starting from $n$ is finite if and only if, for
every neighbor $m$, either the infection does not pass from $n$ to $m$, or
$m$ is infected by $n$, but the outbreak generated by $(n,m)$ is finite.
Hence, defining the following sets of events:
\begin{eqnarray*}
  Z_n &=&\{ z_{nm}=0 , \; \forall m \}, \\
  W_n &=& \cup_m W_{nm},\\ 
  W_{nm} &=& \{ (y_{nm} < \infty ) \cap (z_{nm}=1) \},
\end{eqnarray*}
we can rewrite the condition for $y_n$, expressing the generating function
in Eq.~\eqref{eq_phi} as
\begin{equation}\label{eq_phi_1}
  \phi_n(s)=  E \bigl[ e^{-s [ \prod_m(1- z_{nm})+ \sum_m z_{nm}y_{nm} ]}  |
  Z_{n} \cup W_{n} \bigr]. 
\end{equation}

Since $W_n$ is the union of all independent events $W_{nm}$, we can write
\begin{widetext}
  \begin{eqnarray}
      \label{eq_E}
      E \bigl[ e^{-s [ \prod_m(1- z_{nm})+ \sum_m z_{nm}y_{nm} ]} | Z_{n}
      \cup W_{n} \bigr]P(Z_{n} \cup W_{n}) 
      &=& 
      E \bigl[ e^{-s [ \prod_m(1- z_{nm})+ \sum_m z_{nm}y_{nm} ]} | Z_{n}
      \bigr]P(Z_{n} ) \\
      &+& 
      \sum\limits_m E \bigl[ e^{-s [ \prod_m(1- z_{nm})+ \sum_m z_{nm}y_{nm} ]} | W_{nm}  \bigr]P(W_{nm} ).
\end{eqnarray}
\end{widetext}
The elements in the previous equation are defined as follows:
$P(Z_{n})$ is the probability that node $n$ heals before infecting any
of its $k_n$ neighbors $m$. Since we are dealing with Poisson processes, we
have
\begin{equation}
  P(Z_{n})= \frac{\mu}{\mu + k_n \beta}.
\end{equation}
$P(W_{nm})$ is the probability that $n$ infects $m$, before healing,
conditioned to the fact that the outbreak starting after the first time
step from this pair of infected nodes is finite. We can therefore write it
as
\begin{equation}
  P(W_{nm})= b_{nm}
  \frac{\beta a_{nm}}{\mu + k_n \beta}, 
\end{equation}
so that
\begin{equation}
 P(Z_{n} \cup W_{n}) =
\frac{\mu + \beta \sum_m a_{nm} b_{nm}}{\mu + k_n \beta},
\end{equation}
where $b_{nm}$ is the probability that the outbreak generated by the pair of
adjacent infected nodes $(n,m)$ is finite.  We also have
\begin{equation}
E\bigl[ e^{-s [ \prod_m(1- z_{nm})+ \sum_m z_{nm}y_{nm} ]} | Z_{n} \bigr] = e^{-s}
\end{equation}
and
\begin{eqnarray*}
E \bigl[ e^{-s [ \prod_m(1- z_{nm})+
      \sum_m z_{nm}y_{nm} ]} | W_{nm} \bigr] 
      &=& \phi_{nm}(s) \\
      &\approx&
\phi_n(s)\phi_m(s)
\end{eqnarray*}
where the last factorization derives from the assumption of
independent outbreaks, which, in terms of the sizes, translates into
the assumption that the size of an outbreak generated by a pair $(n,m)$
is the sum of the sizes of the two distinct single-seed outbreaks. In
Sec.~\ref{seq:numeric_synthetic} we present numerical evidence backing up
this factorization.

Denoting $\eta_n = \beta \sum_m a_{nm} b_{nm}$, plugging the previous
equations into Eq.~\eqref{eq_E} and then into Eq.~\eqref{eq_phi_1}, we
obtain
\begin{equation}
  \phi_n(s)= \frac{\mu }{\mu + \eta_n} e^{-s} + \sum\limits_m \frac{\beta
  a_{nm} b_{nm} }{\mu + \eta_n} \phi_n(s)\phi_m(s) .
\end{equation}
Taking the derivative with respect to $s$ and setting $s=0$ we obtain
\begin{eqnarray}
  \phi_n'(0) 
  &=& - \frac{\mu }{\mu + \eta_n} + \phi_n'(0)\sum\limits_m \frac{\beta a_{nm}
  b_{nm} }{\mu + \eta_n} \phi_m(0) \nonumber\\
  &+& \phi_n(0)\sum\limits_m \frac{\beta a_{nm} b_{nm} }{\mu + \eta_n} \phi_m'(0).  
\label{eq_dphi}
\end{eqnarray}
From the definition of the generating function, $\phi_m(0)=1$ while
$\phi_n'(0)$ is given by 
\begin{equation}
  \phi_n'(0)=\frac{d\phi_n(s)}{ds}\bigg|_{s=0}=-\sum\limits_s s P_n(s)=-S_n,
\end{equation}
where $P_n(s)$ is the size distribution of finite outbreaks.  Hence,
Eq.~\eqref{eq_dphi} provides the prediction for the average size of finite
outbreaks
\begin{equation}\label{eq_s_qmf}
S_n=1+ \alpha \sum\limits_m a_{nm} b_n b_m S_m ,
\end{equation}
where we have used the factorization $b_{nm} = b_n b_m$ implied by the
factorization $c_n(t) = c_n(t) c_m(t)$ discussed in
Sec.~\ref{sec:quenched-field-theory}.

We emphasize that the results obtained so far are valid throughout the whole
phase-diagram, since we have not made any assumption on the value of the
parameter $\lambda$.

\section{Numerical results for synthetic networks}
\label{seq:numeric_synthetic}

In this Section we compare the theoretical predictions obtained above with
the results of numerical simulations of SIS dynamics on random networks with
power-law degree distribution $P(k) \sim k^{-\gamma}$. To avoid any form of
correlation by degree we build the networks using the Uncorrelated
Configuration Model (UCM)~\cite{Catanzaro2005}, with minimum degree
$\kmin=3$ and maximum degree $\kmax = \min\{N^{1/2}, N^{1/(\gamma-1)}\}$.
We consider in particular two values of the exponent $\gamma$, corresponding
to different properties of the topology and of the SIS dynamics taking place
on it.

Considering the degree exponent $\gamma=2.25$ as representative of the case
$\gamma<5/2$, we have highly heterogeneous networks with largest
eigenvalue $\Lambda_M$ well approximated by the ratio
$\av{k^2}/\av{k}$~\cite{Chung2003}.  The corresponding principal eigenvector
is localized on a subextensive subgraph coinciding with the set of nodes
with largest core index in the K-core decomposition~\cite{PastorSatorras2016}.
For these topologies, quenched mean-field theory and annealed network theory
give, in the large $N$ limit, the same critical properties, that agree very
well with numerical simulations~\cite{Ferreira2012}.  Based on this, we
expect the present approach to be successful in predicting the spreading
influence of individual nodes, the agreement improving as larger networks
are considered.

The value $\gamma=3.5$, as an instance of networks with $\gamma>5/2$, shows
instead markedly different spectral properties~\cite{Chung2003}.  The
principal eigenvector is in this case localized on the hub with largest
degree and on its immediate neighbors~\cite{Goltsev2012}.
The corresponding eigenvalue is
approximately given by $1/\sqrt{\kmax}$. In this case quenched mean-field
predictions are very different from those of the annealed network theory.
Neither of the theories agrees well with numerical simulations. In
particular the quenched mean-field theory estimate for the threshold
provides only a lower-bound for the true value, which is
determined by a very complex interplay between hubs in the network, which
are distant but can mutually reinfect each other~\cite{Castellano2020}.
Therefore we do not expect a full agreement between theory and simulations,
starting from the position of the threshold, which is expected to be larger
than $\lambda=1$ and to grow further as network size is increased.

We simulate SIS dynamics by means of an optimized Gillespie
algorithm~\cite{Cota2017}. Network size is generally $N=10^4$ nodes.
Average values are obtained by performing at least 1000 realizations of the
stochastic process for each seed.  The numerical evaluation of the
observables we are interested in is intrinsically difficult, because the
distinction between finite and infinite outbreaks is clear-cut only for
infinite networks. For networks of finite size all outbreaks last
necessarily only a finite amount of time, reaching eventually the absorbing,
healthy state. It is nevertheless possible to distinguish between truly
finite outbreaks and putatively infinite ones, which end only because of the
network finite size, by identifying two different components in the
distribution of outbreak durations.  See Appendix~\ref{choice} for details.
Close to the threshold, the distinction becomes conceptually impossible, as
the two components get inextricably superposed. This makes a comparison
between theory and simulations unfeasible in the vicinity of the critical
point.

\begin{figure}
  \includegraphics[width=\columnwidth]{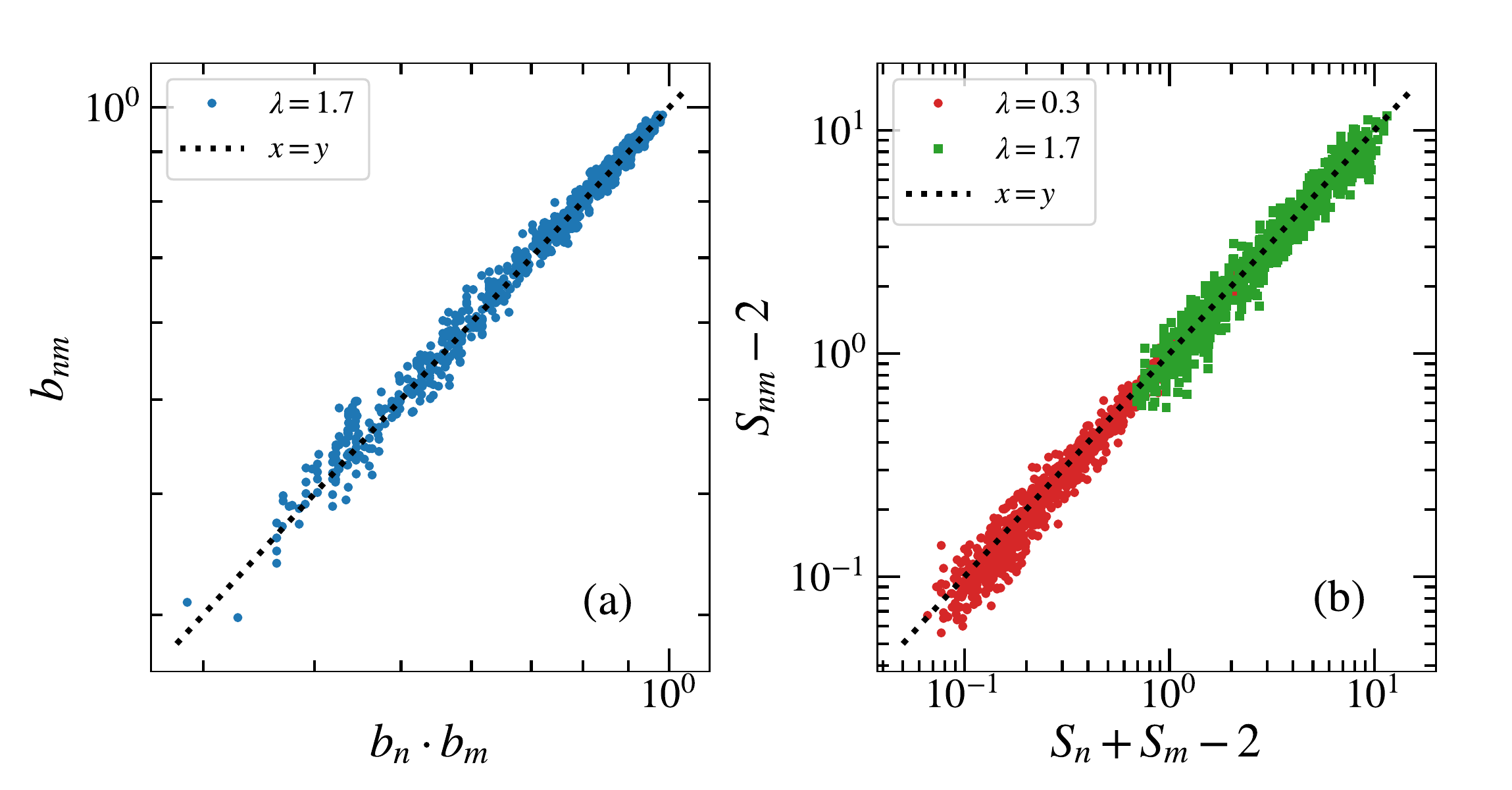}

  \caption{Numerical check of assumptions used in our
    theoretical analysis. (a) Factorization of the probability that an
    outbreak initiated at two randomly chosen adjacent nodes $(n,m)$ is
    finite. (b) Assumption about the average size of finite outbreaks of
    an outbreak initiated at two randomly chosen adjacent nodes.
    Results correspond to networks with
    degree exponent $\gamma=2.25$ and size $N=10^4$.}
  \label{fig:factorization}
\end{figure}

As a preliminary step, we have checked in numerical simulations the validity
of the factorization for the probability $c_{nm}(t)$ that an outbreak
starting from two adjacent seed nodes $(n,m)$ lasts a time smaller than or
equal to $t$. To arrive at Eq.~\eqref{eq_c} it was assumed the factorization
$c_{nm}(t) = c_n(t) c_m(t)$.
Considering the limit of
infinite time, this factorization implies that $b_{nm} = b_n b_m$, i.e.,
the probability of an outbreak starting with a pair of nodes being
finite is equal to the product of the probabilities that each node induces
independently a finite outbreak. The validity of this factorization is
checked numerically in Fig.~\ref{fig:factorization}(a). Additionally, in the
calculation of the average outbreak size, a second assumption was made,
namely that the average size $S_{nm}$ of a finite outbreak starting from a
pair of infected nodes $(n,m)$ is equal to sum of the average sizes of
finite outbreaks starting independently from $n$ and $m$, namely,
$S_{nm} = S_n + S_m$.
This assumption is numerically checked in
Fig.~\ref{fig:factorization}(b). These two results support the validity of
the theoretical approach developed in the previous Section.

\subsection{$\gamma = 2.25$}
\begin{figure}
  \includegraphics[width=\columnwidth]{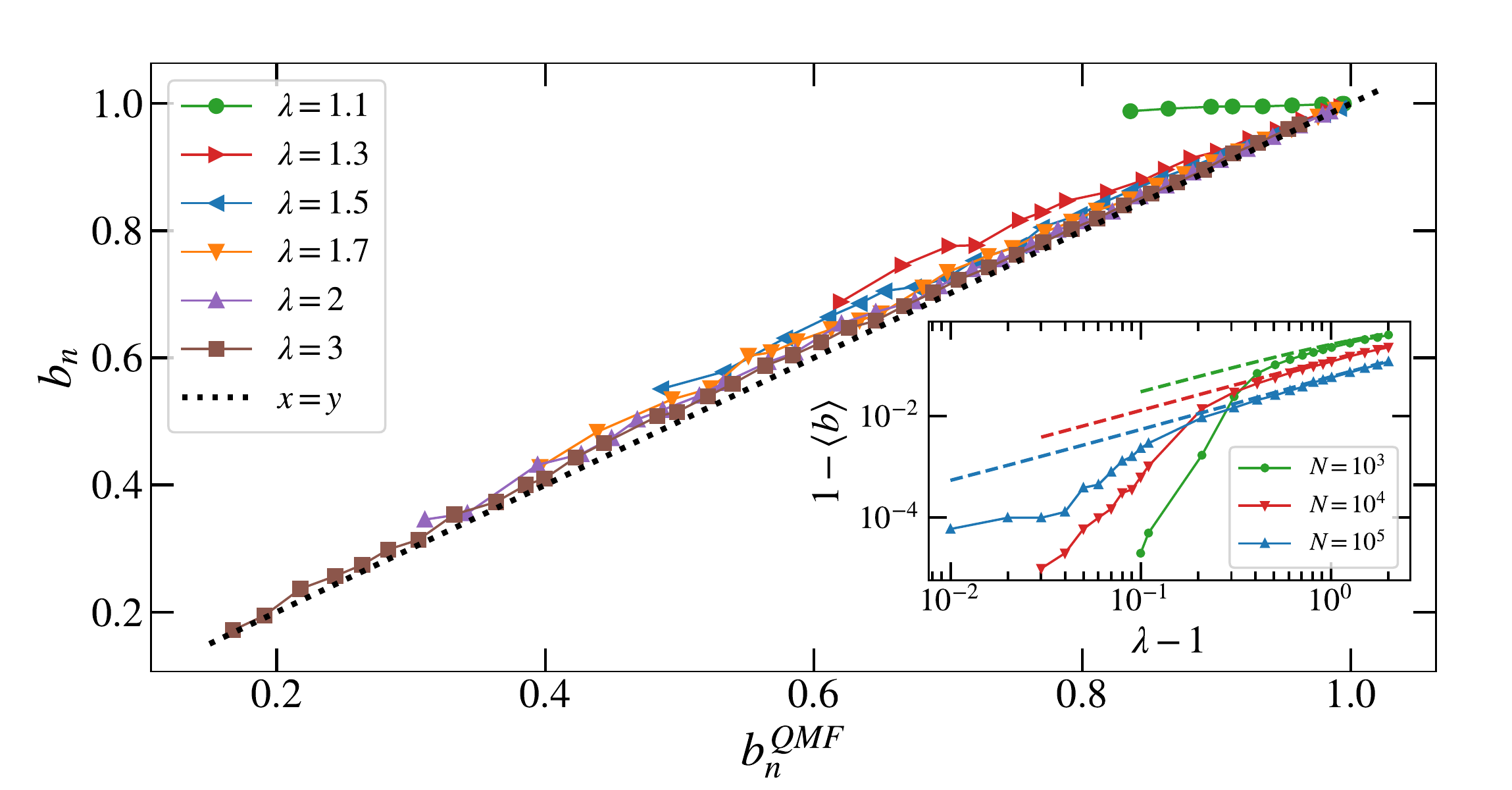} 
  \caption{Numerical results for the probability $b_n$ in networks with
    degree exponent $\gamma=2.25$ and size $N=10^4$. Main plot: Comparison
    of $b_n$ vs the theoretical prediction $b_n^{QMF}$ obtained from
    Eq.~\eqref{eq_b_qmf} for different values of $\lambda = \alpha
  \Lambda_M$. Inset: Comparison of the average $1-\av{b}$ vs the theoretical
prediction from Eq.~\eqref{eq_b_qmf} as a function of $\lambda$. } 
  \label{bnvsbnqmf}
\end{figure}

Figure~\ref{bnvsbnqmf} compares the probability $b_n$ to observe a finite
outbreak starting from seed $n$, measured in simulations, with the
predicted value given by Eq.~\eqref{eq_b_qmf}.  The agreement is excellent,
except for the smallest value of $\lambda$.  This discrepancy is a
consequence of the fact that the effective threshold for finite size is
larger than its asymptotic value $\lambda_c=1$: hence for $\lambda=1.1$ the
theory predicts $b_n<1$ while for most seeds $b_n=1$, as the system is below
the effective threshold. This interpretation is confirmed by the inset,
where the average value $\av{b} = N^{-1} \sum_n b_n$ is plotted against
$\lambda$: $\av{b}$ follows well the theoretical prediction, valid in the
thermodynamic limit, only for $\lambda-1 > 0.2$.

\begin{figure}
  \includegraphics[width=\columnwidth]{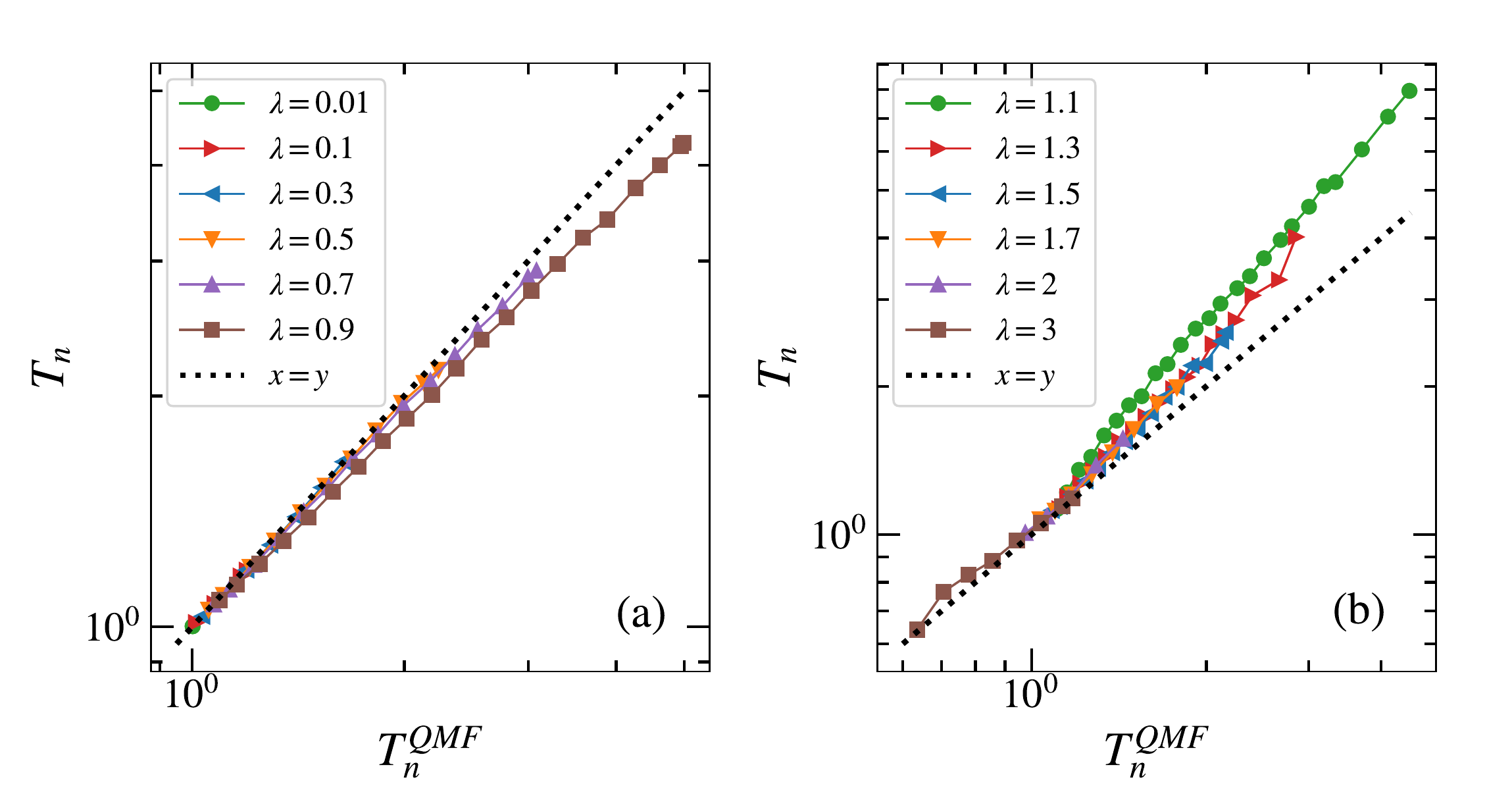}

  \caption{Comparison of the numerically evaluated average duration of
    finite outbreaks $T_n$ vs the theoretical prediction $T_n^{QMF}$
    obtained from Eqs.~\eqref{eq_df} and~\eqref{eq_T_qmf} in networks with
    degree exponent $\gamma=2.25$ and size $N=10^4$. (a) Theoretical
    subcritical regime $\lambda<1$. (b) Theoretical supercritical regime
  $\lambda>1$.}
 
  \label{TnvsTnqmf}
\end{figure}

\begin{figure}
  \includegraphics[width=\columnwidth]{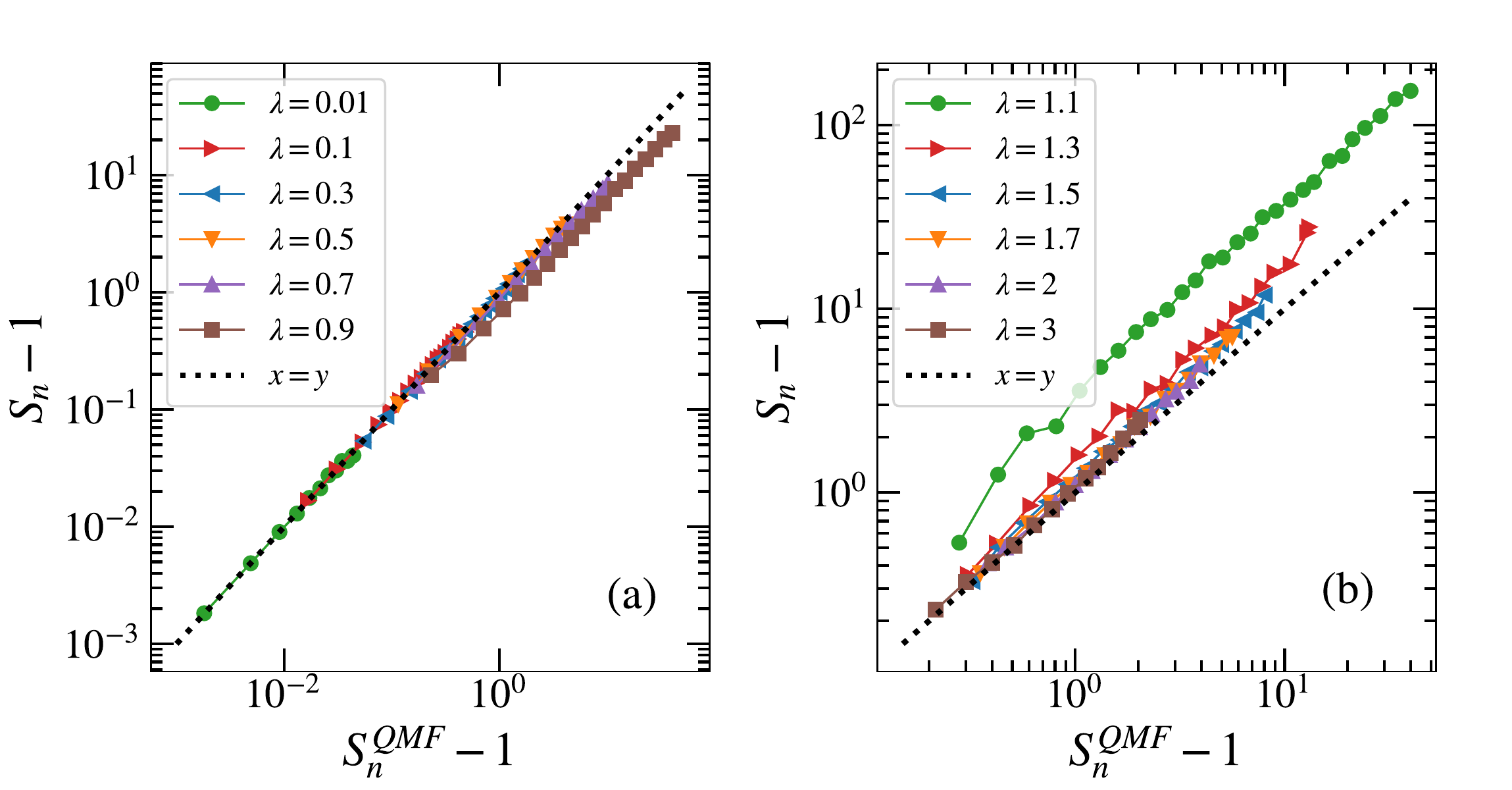}
\caption{Comparison of the numerically evaluated average size of
    finite outbreaks $S_n$ vs the theoretical prediction $S_n^{QMF}$
    (from Eq.~\eqref{eq_s_qmf}) in networks with
    degree exponent $\gamma=2.25$ and size $N=10^4$. (a) Theoretical
    subcritical regime $\lambda<1$. (b) Theoretical supercritical regime
  $\lambda>1$.}
 \label{SnvsSnqmf}
\end{figure}

In Fig.~\ref{TnvsTnqmf} we report the comparison between simulations and
theory concerning the duration of finite outbreaks, for both subcritical
and supercritical values of the parameter $\lambda$. Also in this case
the agreement is fully satisfactory, the only limited discrepancies occurring
around the transition, as expected.
A very good agreement is found also in Fig.~\ref{SnvsSnqmf}, where the average size
of finite outbreaks starting in node $n$ is compared with the solution of
Eq.~\eqref{eq_s_qmf}.
We conclude that the theoretical approach presented above describes very
accurately the spreading influence of nodes in random uncorrelated networks
with $\gamma<5/2$.

\begin{figure}
  \includegraphics[width=\columnwidth]{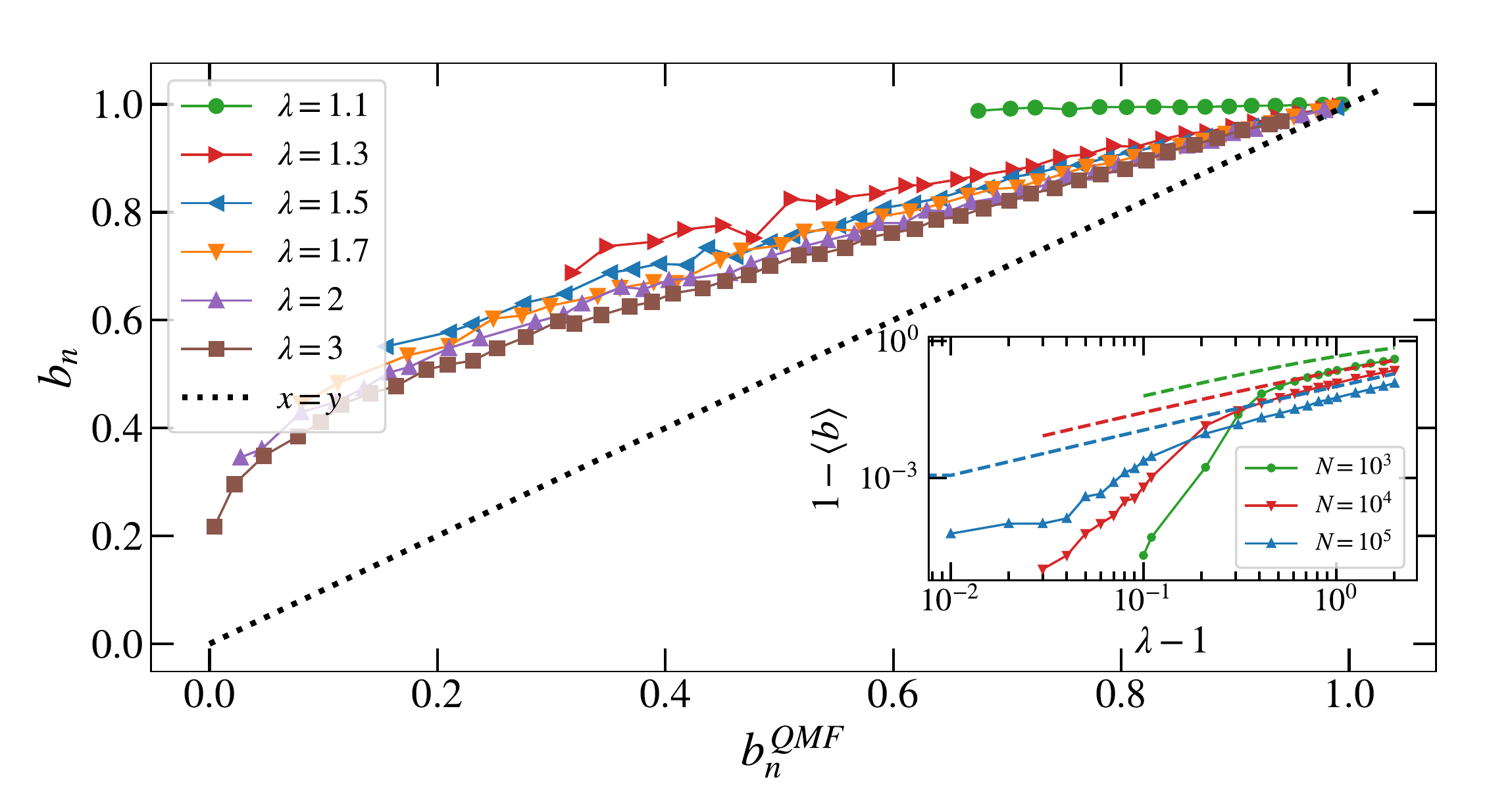} 
  \caption{Numerical results for the probability $b_n$ in networks
      with
    degree exponent $\gamma=2.25$ and size $N=10^4$. Main plot: Comparison
    of $b_n$ vs the theoretical prediction $b_n^{QMF}$ obtained from
    the discrete time theory~\cite{PouxMedard2020} for different values
    of $\lambda = \alpha \Lambda_M$.
    Inset: Comparison of the average $1-\av{b}$ vs the theoretical
    prediction for discrete time~\cite{PouxMedard2020} as a function
    of $\lambda$.}
  \label{discrete}
\end{figure}

One might wonder whether a similar good agreement between
  theory and simulations could have been achieved by using the theory
  for discrete time SIS dynamics presented in
  Ref.~\cite{PouxMedard2020}. We notice that the result for the
  average size of finite outbreaks is identical in continuous and
  discrete time in the subcritical regime; however, this is a
  coincidence that does not happen for the other observables and
  in the supercritical region.
  As an example,
  Fig.~\ref{discrete} shows the strong difference between the
  numerical probability $b_n$ to observe a finite avalanche and the
  corresponding discrete time prediction. This plot confirms that
  taking into account the continuous time nature of the dynamics is
  necessary for correctly predicting the spreading influence.

\subsection{$\gamma = 3.5$}

\begin{figure}
  \includegraphics[width=\columnwidth]{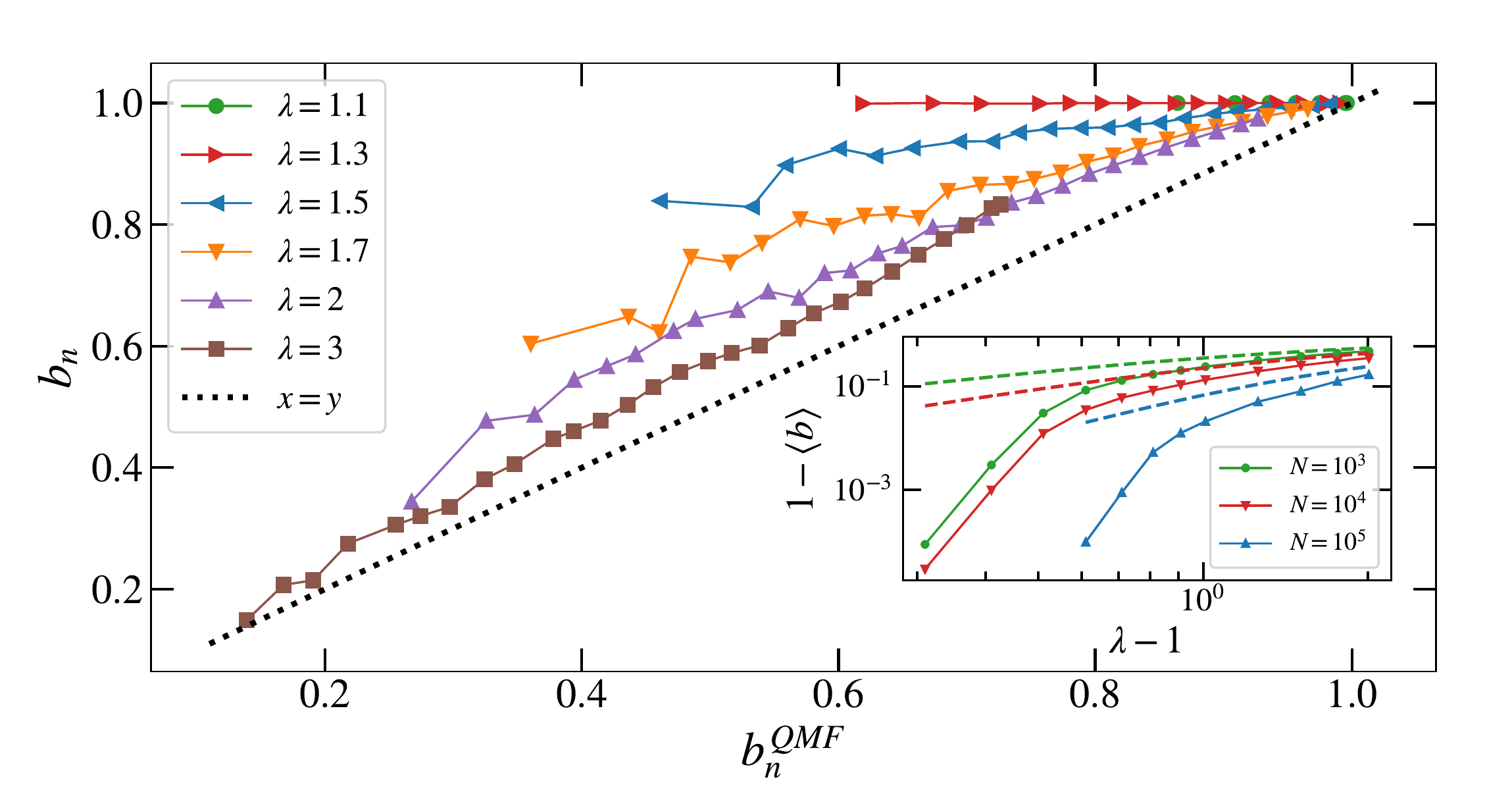} 

  \caption{Numerical results for the probability $b_n$ in networks with
    degree exponent $\gamma=3.50$ and size $N=10^4$. Main plot: Comparison
    of $b_n$ vs the theoretical prediction $b_n^{QMF}$ obtained from
    Eq.~\eqref{eq_b_qmf} for different values of $\lambda = \alpha
  \Lambda_M$. Inset: Comparison of the average $1-\av{b}$ vs the theoretical
prediction from Eq.~\eqref{eq_b_qmf} as a function of $\lambda$.}
 \label{bnvsbnqmf2}
\end{figure}

\begin{figure}
  \includegraphics[width=\columnwidth]{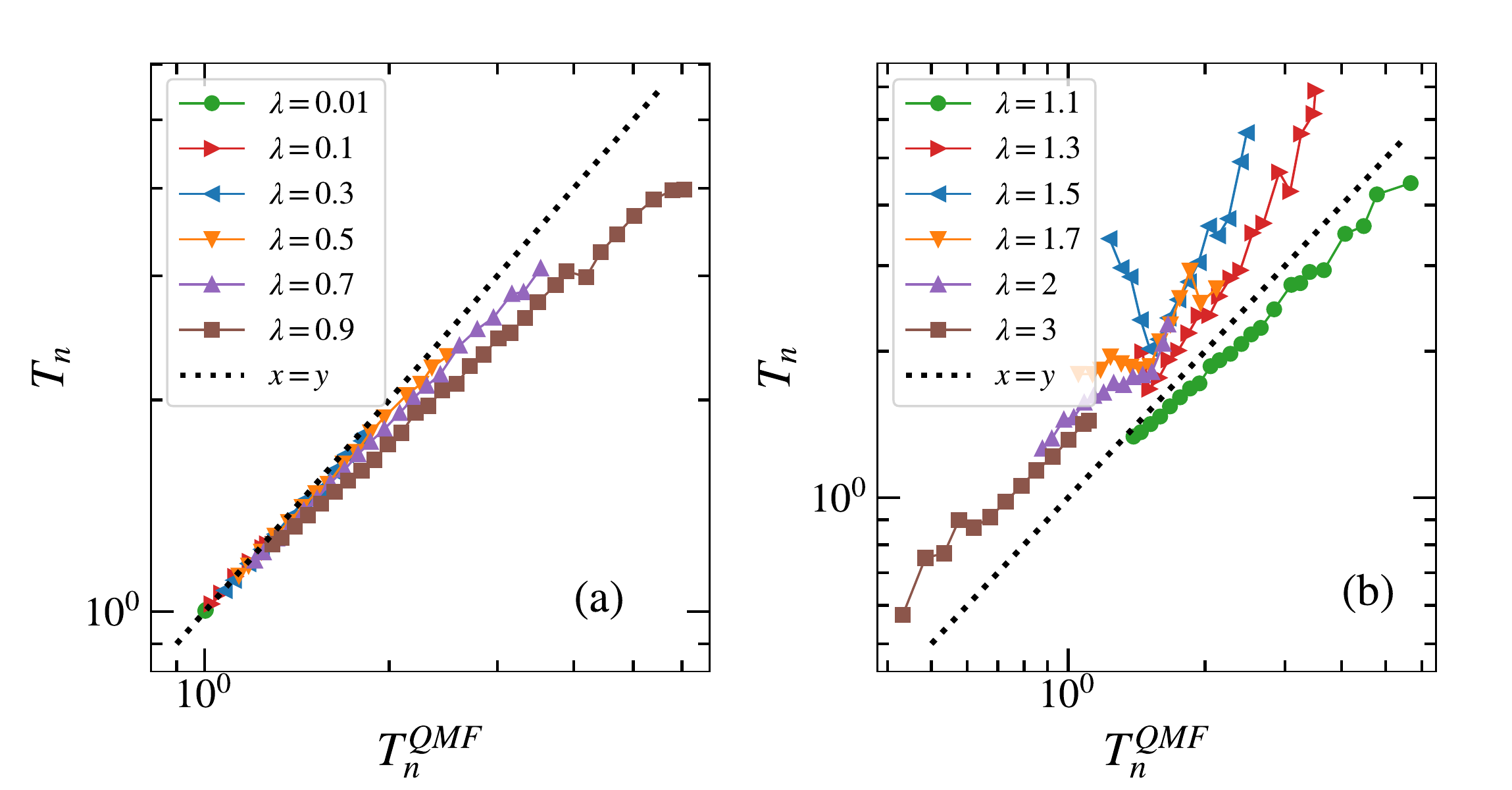}
  \caption{Comparison of the numerically evaluated average duration of
    finite outbreaks $T_n$ vs the theoretical prediction $T_n^{QMF}$
    evaluated from Eqs.~\eqref{eq_df} and~\eqref{eq_T_qmf} in networks with
  degree exponent $\gamma=3.50$ and size $N=10^4$. (a) Theoretical
subcritical regime $\lambda<1$. (b) Theoretical supercritical regime
$\lambda>1$.}
 \label{TnvsTnqmf2}
\end{figure}

\begin{figure}
  \includegraphics[width=\columnwidth]{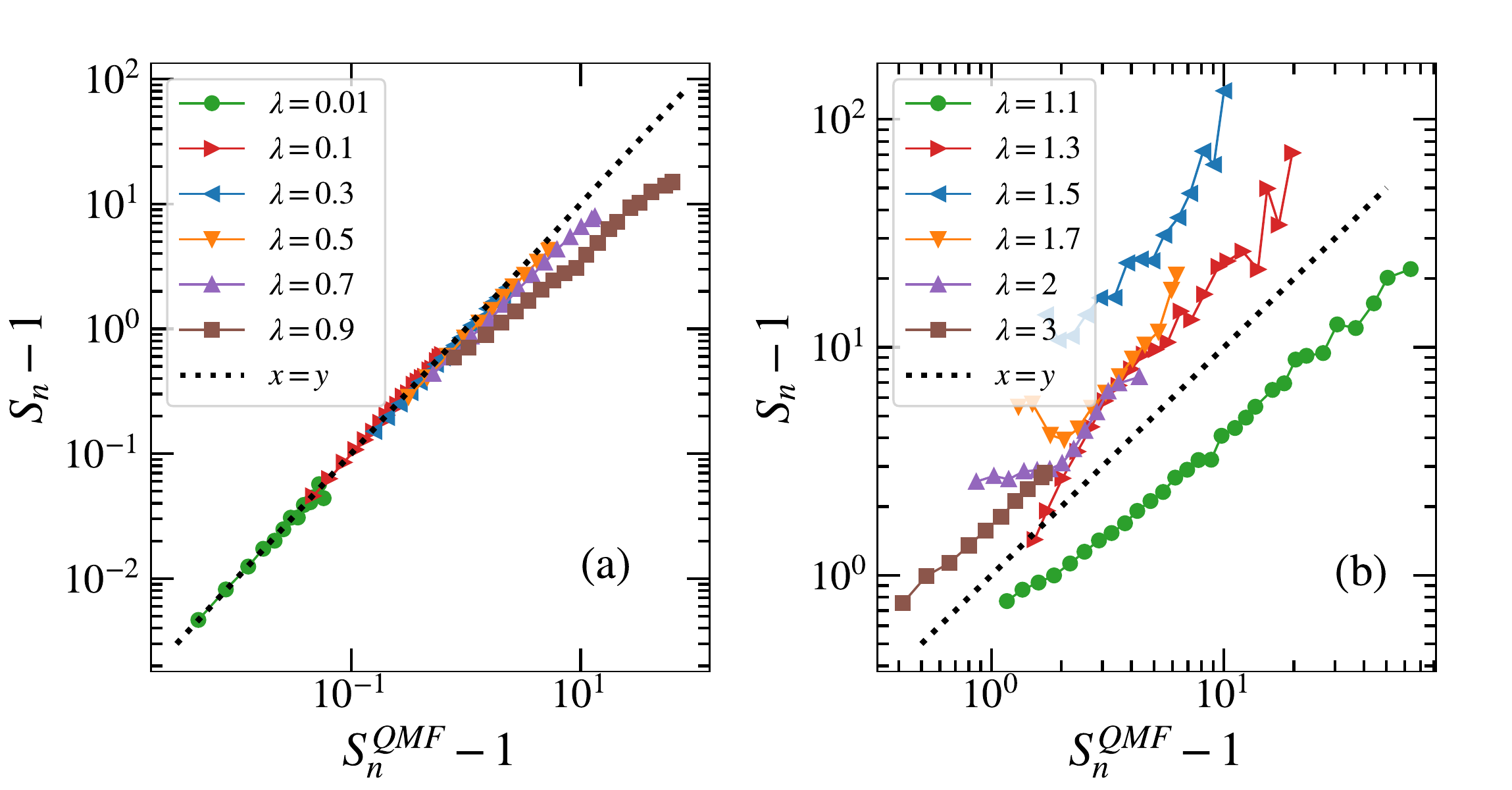}
  \caption{Comparison of the numerically evaluated average size of finite
    outbreaks $S_n$ vs the theoretical prediction $S_n^{QMF}$ evaluated
    from Eq.~\eqref{eq_s_qmf} in networks with degree exponent $\gamma=3.50$
    and size $N=10^4$. (a) Theoretical subcritical regime $\lambda<1$. (b)
  Theoretical supercritical regime $\lambda>1$.}
 
  \label{SnvsSnqmf2}
\end{figure}

As mentioned above, for $\gamma=3.5$ we do not expect a perfect agreement
between theory and simulations, because of the known shortcomings of QMF
theory for these mildly heterogeneous networks.  This is confirmed by
Fig.~\ref{bnvsbnqmf2}. The probability to originate a finite outbreak
starting from node $n$ is definitely larger than the theoretical prediction
given by Eq.~\eqref{eq_b_qmf}. Only for strongly supercritical cases
($\lambda \geq 3$) the discrepancy becomes small. This is a consequence of
the fact that the effective threshold in simulations is much larger than
$\lambda_c=1$ (see the inset).  At variance with the $\gamma<5/2$ case, the
disagreement becomes even larger as $\lambda$ grows.
The qualitative difference with the case $\gamma=2.25$ discussed above is
apparent also in the comparison between theoretical and numerical results
for the average duration of finite outbreaks, see Fig.~\ref{TnvsTnqmf2}.
While for strongly subcritical values of $\lambda$ the agreement is
reasonably good, the performance of the theory is reduced close to
$\lambda_c$ and in a large interval of $\lambda$ values above it.
This is even more evident when finite outbreak average sizes are considered,
see Fig.~\ref{SnvsSnqmf2}.

Although it is clear that QMF theory implies a systematic miscalculation of
the spreading influence in this regime of $\gamma$ values, it is remarkable
that for networks of this size errors are not exceedingly large.  While we
know that for larger systems the inaccuracy would be larger, still the
theory can be taken as a fair approximation for not too small networks.

\begin{figure}
  \includegraphics[width=\columnwidth]{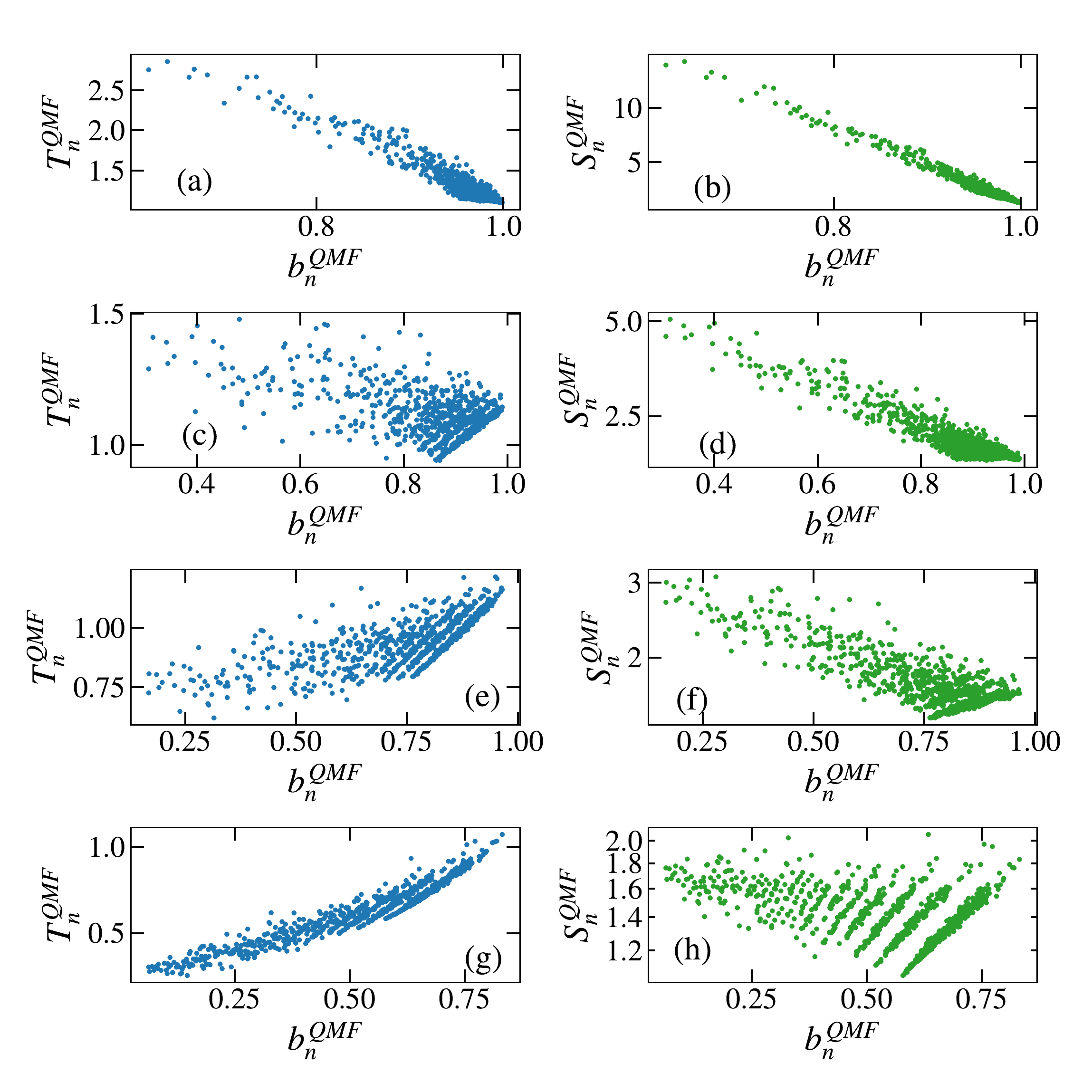}
  \caption{Theoretically predicted average duration (left) and size (right)
    of finite outbreaks vs probability of having a finite outbreak $b_n$
    in networks with degree exponent $\gamma=2.25$ and size $N=10^4$.
    (a) and (b):$\lambda=1.3$; (c) and (d):$\lambda=2$; (e) and
    (f):$\lambda=3$;  (g) and (h):$\lambda=6$}
  \label{vsbn}
\end{figure}

We conclude this section by answering to a question that naturally arises
due to the plurality of observables quantifying spreading influence for SIS:
Do all definitions identify the same nodes as more influentials?
In Fig.~\ref{vsbn} we plot the duration $T_n$ and size $S_n$ as a function
of the probability $b_n$ to have a finite outbreak, for $\gamma=2.25$
and various values of $\lambda>1$.
We see that for small $\lambda$, $T_n$ and $S_n$ decrease with $b_n$, as
expected. In this case a node originating many infinite outbreaks is also a
good spreader generating large finite outbreaks.  Rather surprisingly,
things change for larger $\lambda$.  In such a case, the nodes having high
probability of giving rise to infinite outbreaks, generate {\em shorter}
finite outbreaks.
Hence they are good spreaders in one
sense and bad ones in the other.  Notice however that outbreaks in this case
are minuscule.

\subsection{Centralities as predictors}

In the previous subsection we have shown that the QMF theoretical approach
provides good predictions for the spreading influence of individual nodes in
uncorrelated random networks.  Hence we have a way to calculate the size of
a finite outbreak or its duration, with reasonable accuracy, without actually
performing simulations.  Another relevant issue in this context has
to do with the correlation between spreading influence and network
centralities~\cite{Kitsak2010,Lu2016,Chen2012,Li2014,Malliaros:2016aa,Chen:2019aa}.
Is the spreading influence of a given node predictable based on some
topological property, such as degree, eigenvalue centrality or the many
other centralities available on the market?  We investigate this issue for
the synthetic networks considered above, focusing on degree, eigenvector and
K-core centralities, taking advantage of the knowledge
about SIS dynamics in random networks gained in recent
years.

\begin{figure}
  \includegraphics[width=\columnwidth]{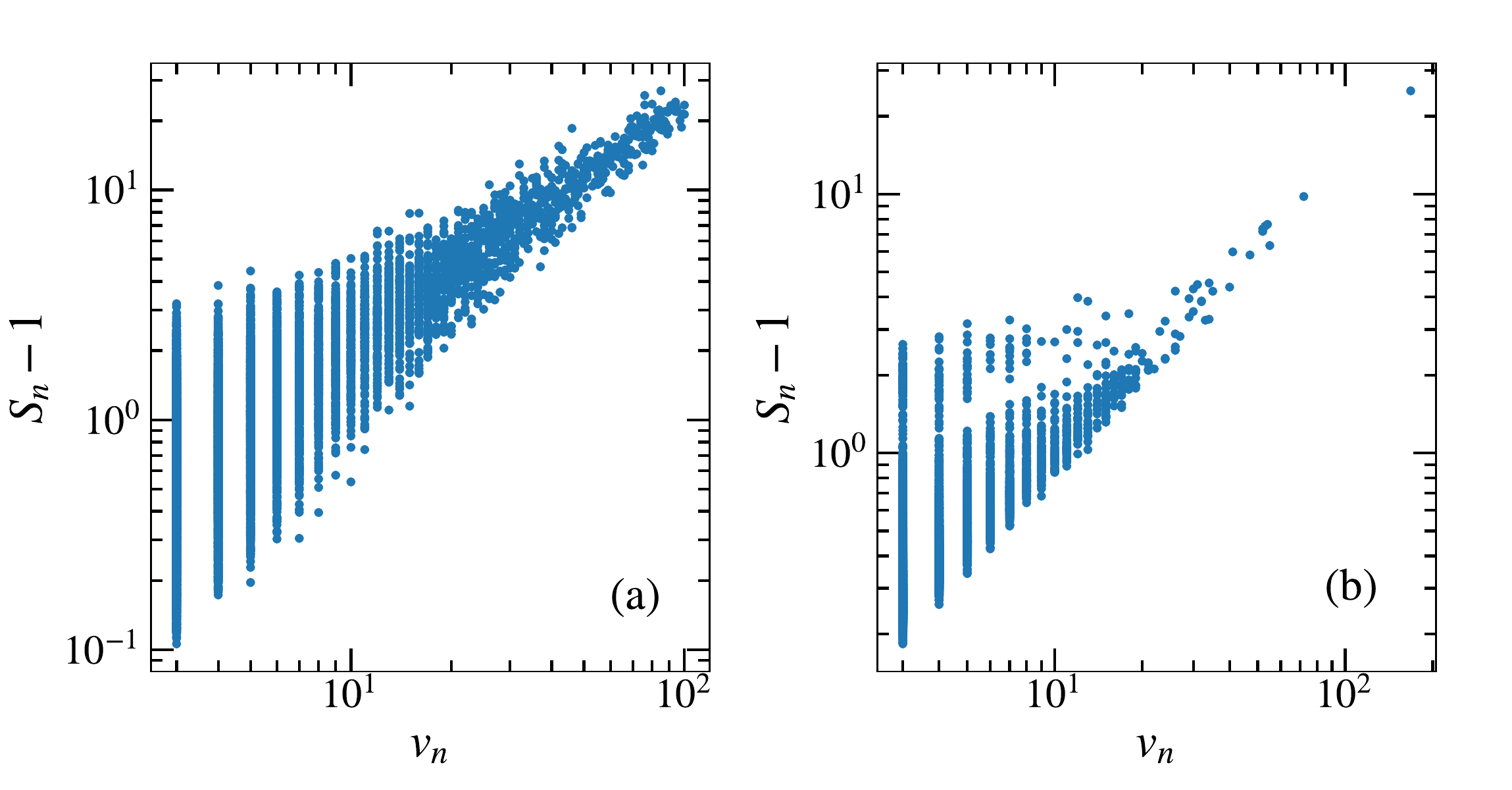}
 
  \caption{Comparison of the numerically evaluated average size of finite
    outbreaks $S_n$ vs the degree centrality $k_n$ in networks of size
    $N=10^4$. (a) Degree exponent $\gamma=2.25$. (b) Degree exponent
    $\gamma=3.50$. Simulations correspond to $\lambda = 0.9$. } 
    \label{Svsk}
\end{figure}

In Fig.~\ref{Svsk} we plot the average finite outbreak size as a function
of the degree $k_n$ of the seed node. While there is clearly a strong correlation
between the two quantities, it is evident that the correlation is far from
perfect: some nodes with degree equal to the minimum $\kmin$ generate on
average outbreaks larger than some nodes with degree even 10 times larger.
This clearly shows that the annealed network assumption that degree
completely determines the spreading properties of each node is only a rough
approximation.

\begin{figure}
  \includegraphics[width=\columnwidth]{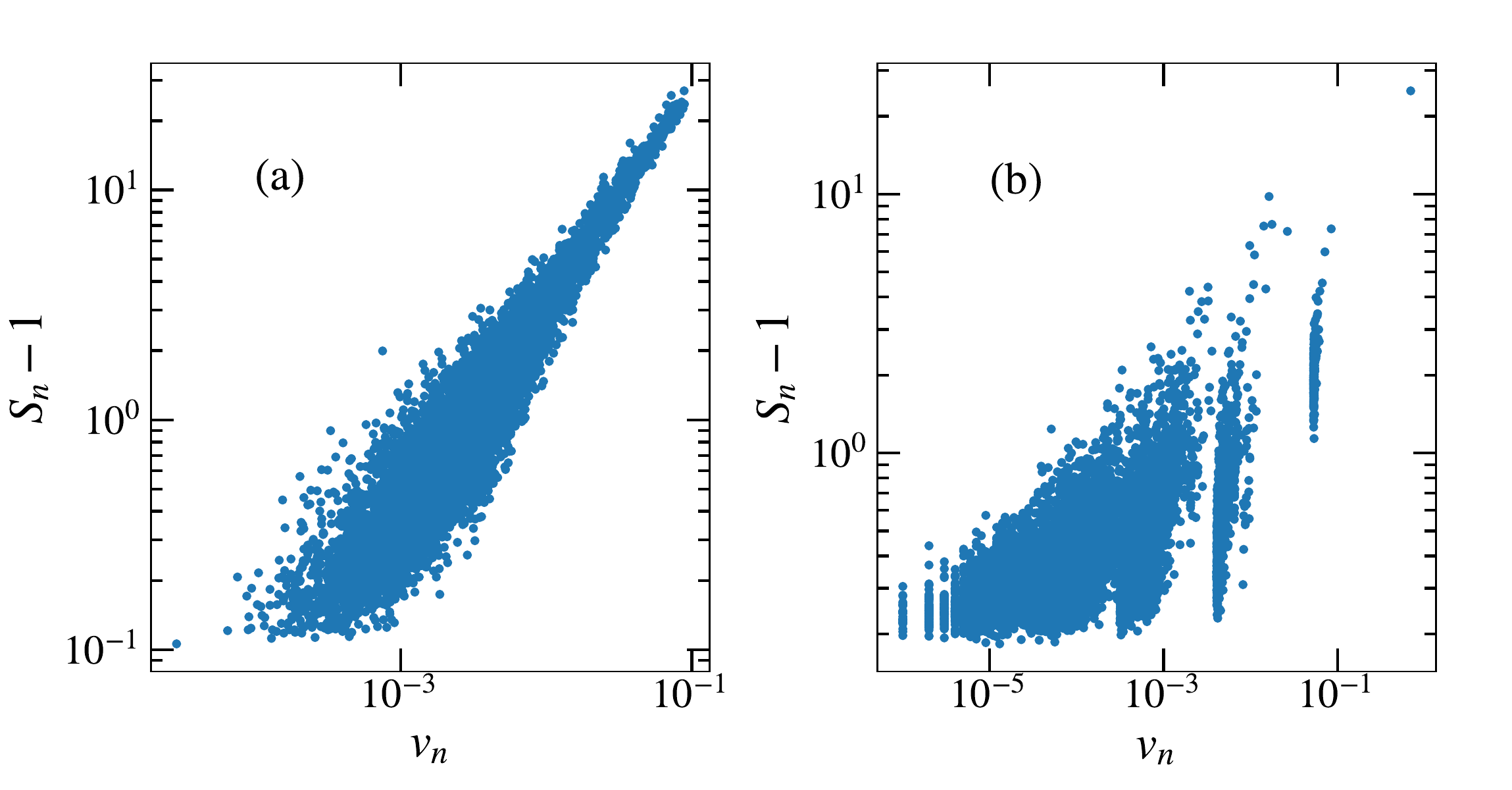}

  \caption{Comparison of the numerically evaluated average size of finite
    outbreaks $S_n$ vs the eigenvector centrality $\nu_n$ in networks of size
    $N=10^4$. (a) Degree exponent $\gamma=2.25$. (b) Degree exponent
    $\gamma=3.50$. Simulations correspond to $\lambda = 0.9$. Networks used
  are the same as in Fig.~\ref{Svsk}.}
 
  \label{SvsPEV}
\end{figure}

Fig.~\ref{SvsPEV} shows instead the same  $S_n$ values as a function of the
eigenvector (EV) centrality $\nu_n$, defined as the component on node
$n$ of the principal eigenvector of the adjacency matrix~\cite{Bonacich72}.
From this plot it appears that, in the case $\gamma=2.25$, the EV
centrality is a better predictor of spreading influence than degree.
We can estimate quantitatively the accuracy of both predictions by calculating
the linear correlation coefficient $R$ between $S_n$ and the corresponding
centrality. The values $R = 0.952$, found for degree, and $R=0.982$, for EV
centrality, indicate the superior performance of the latter.

For $\gamma=3.5$ EV centrality is still correlated with $S_n$,
but the presence of some structure clearly emerges.  Actually, this plot can be
interpreted based on the physical picture of the SIS transition sketched
above. For $\gamma=3.5$ the principal eigenvector is localized around the
largest hub in the network and its components decay as a function of the
distance from it~\cite{Goltsev2012}. This explains the vertical bands
occurring in the right side of the plot, corresponding to nodes at distance
1, 2, 3 from the hub~\footnote{For $\gamma=3.50$ the effect of the distance
from the hub
  is more visible if a large gap exists between the degree of the first and of
  the second hub. For this reason we selected a suitable realization of the network
  with such a large gap.}.

\begin{figure}
  \includegraphics[width=\columnwidth]{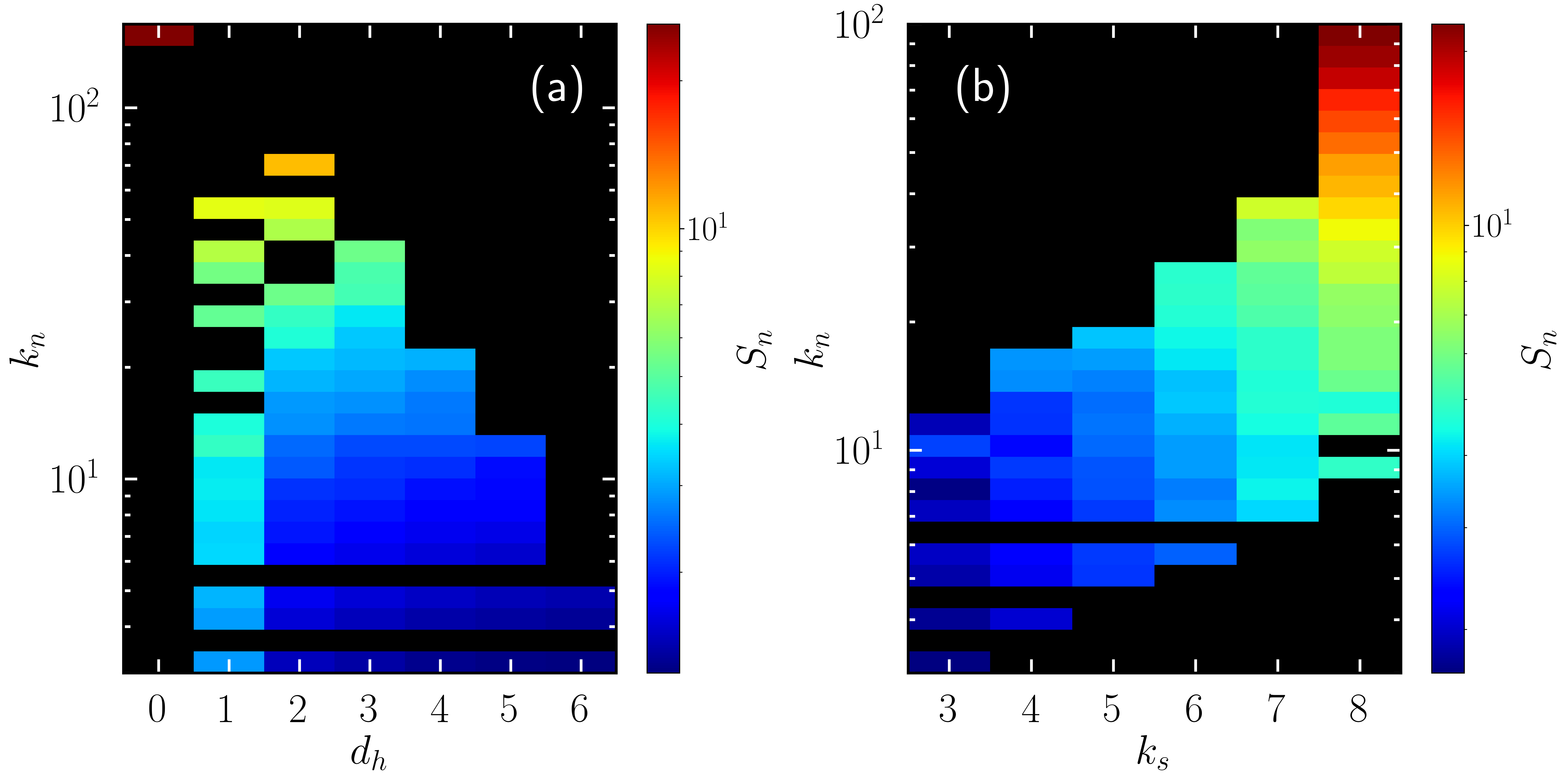}
  \caption{Dependence of the average size $S_n$ of finite outbreaks
    generated as a function of the centrality properties of the seed node.
    (a) $S_n$ as a function of the degree $k_n$ and distance from the hub
    $d_h$ in a network of degree exponent $\gamma = 3.50$. (b)  $S_n$ as a
    function of the degree $k_n$ and the coreness $k_s$, defined as the
    K-core index of the node in the K-core decomposition, in a network of
  degree exponent $\gamma = 2.25$. Simulations performed at $\lambda = 0.9$.
Networks used are the same as in Fig.~\ref{Svsk}. }
 \label{Svsd}
\end{figure}

A node with $k=\kmin$ at distance 1 from the hub is a
much better spreader than a very distant node, even if much more connected.
This is very clearly seen if we plot, as in Ref.~\cite{Kitsak2010}, $S_n$ as
a function of the degree and of the distance from the largest hub in the
network, see Fig.~\ref{Svsd}(a).
\begin{figure}
  \includegraphics[width=\columnwidth]{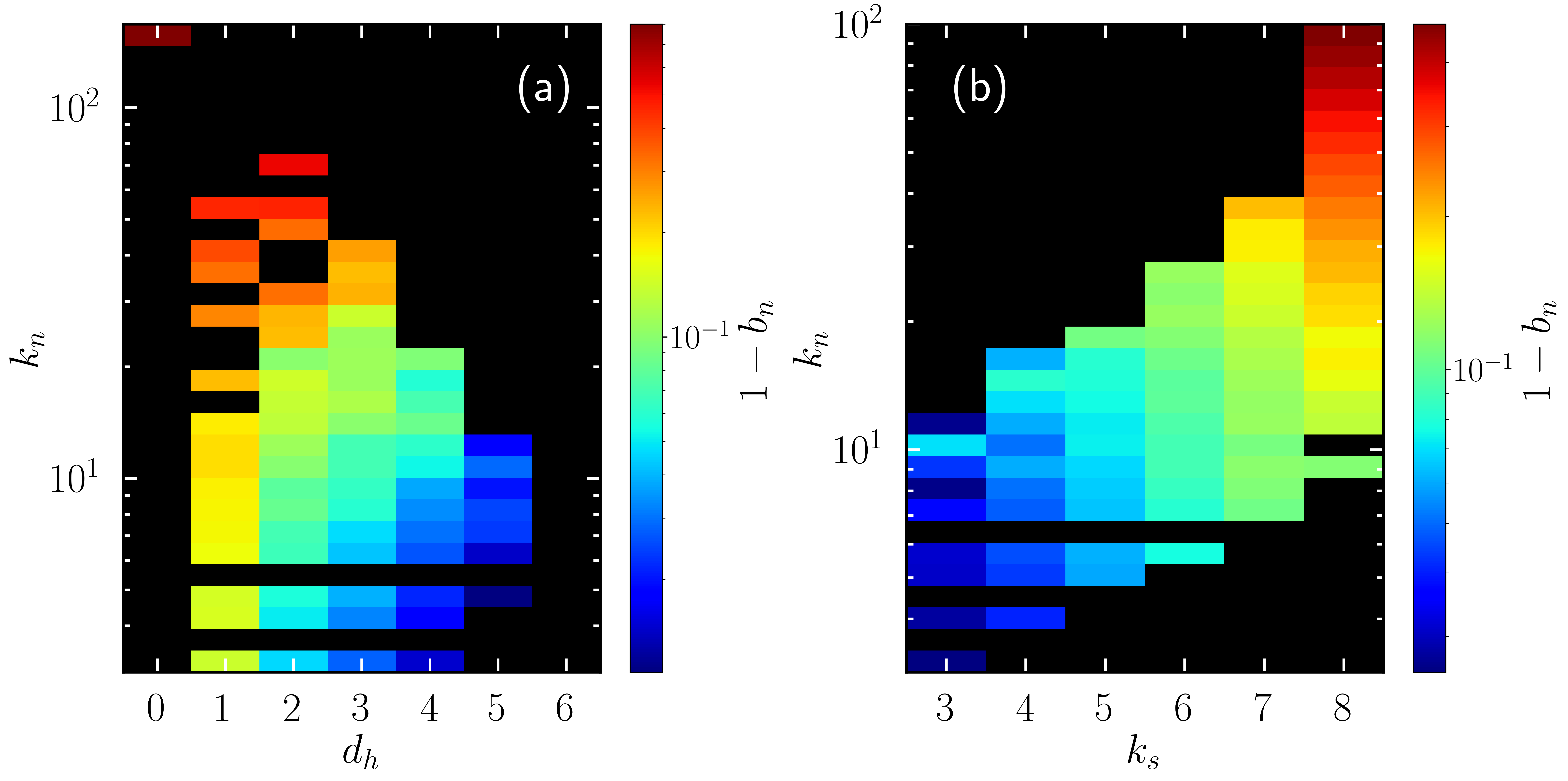}
  \caption{Dependence of the probability $1-b_n$ to generate
      an infinite outbreak (i.e. a steady state) as a function of the
      centrality properties of the seed node.  (a) $1-b_n$ as a
      function of the degree $k_n$ and distance from the hub $d_h$ in
      a network of degree exponent $\gamma = 3.50$. (b) $1-b_n$ as a
      function of the degree $k_n$ and the coreness $k_s$ in a network
      of degree exponent $\gamma=2.25$. Network and
      simulation parameters as in Fig.~\ref{Svsd}.}
 \label{Bnvsd}
\end{figure}
It is clear that, while the spreading influence depends on the degree,
there is also a clear dependence on the distance from the network
hub. Nodes with the same degree are much better spreaders if they are
close to the node with highest degree.  For $\gamma<5/2$ we know
instead that a central role in SIS dynamics is played by the mutually
interconnected subgraph identified by the maximum core index in the
K-core decomposition. Figure~\ref{Svsd}(b) nicely confirms this
interpretation. In Fig.~\ref{Bnvsd} we present an analogous
  analysis, performed in this case on the probability $1-b_n$ to
  generate an infinite outbreak. The correlations with $k_n$ and $k_s$
  are now slightly weaker than for the average finite outbreak size,
  but the figure still shows that these centralities are rather good
  predictors for the emergence of a steady state from an infection
  seeded in a single node.

\section{Real-world networks}

So far we have considered random uncorrelated synthetic networks, whose
topological properties are well controlled and suitable for performing
analytical calculations, but far from those found in the real-world, where
correlations, short loops, communities and other mesostructures
abound~\cite{Newman10}.  The theory developed above can
nevertheless be applied to any type of network. Is it able to accurately
predict $b_n$, $T_n$ and $S_n$ for SIS dynamics on real-world structures?
If not, what are the topological features that invalidate it?

\begin{figure}
  \includegraphics[width=\columnwidth]{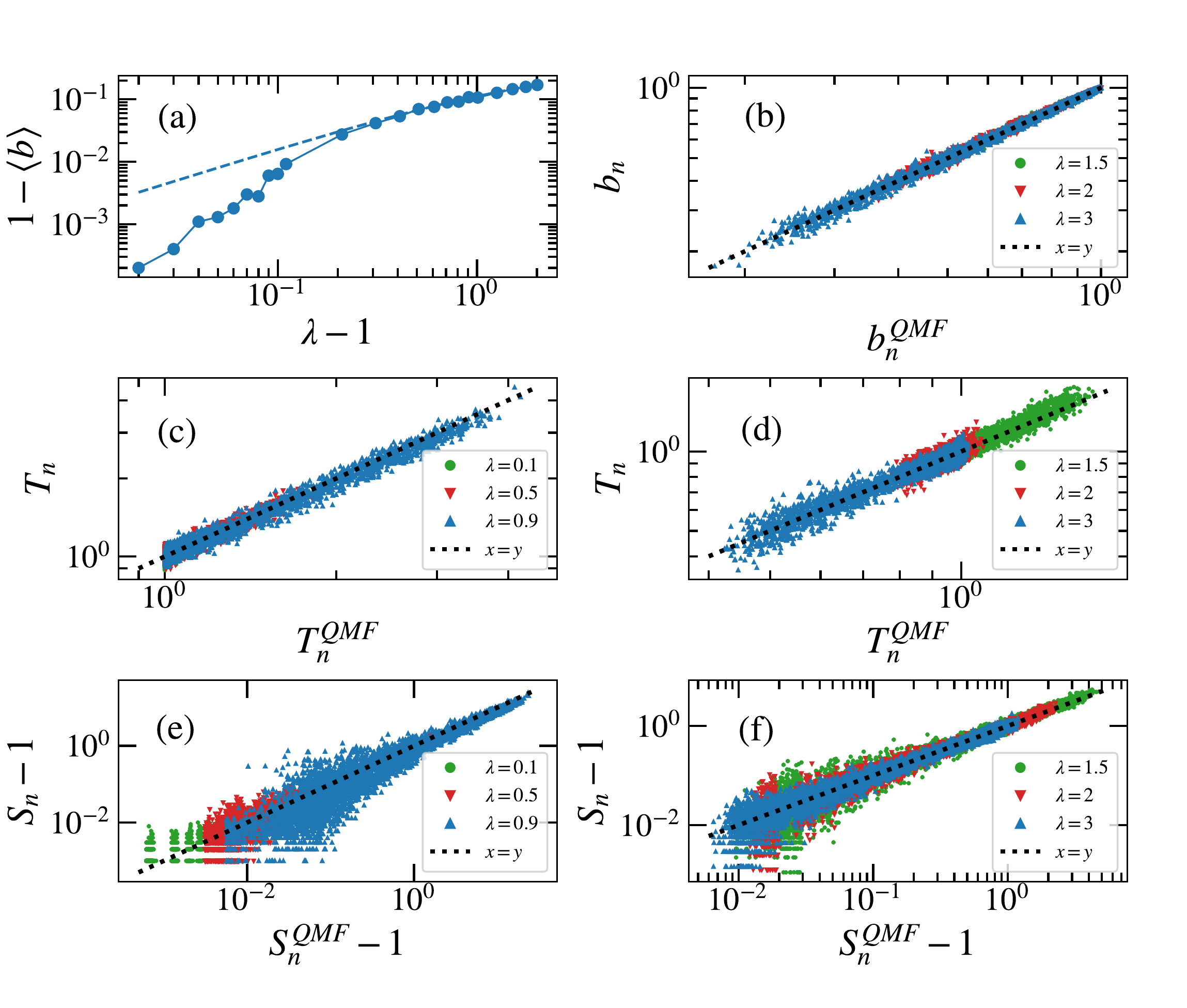}
  \caption{Numerical results for the Tennis real network. (a) $1-\av{b}$ as
    a function of $\lambda$; (b) Numerical $b_n$ vs the theoretical
    prediction; (c) and (d) Average outbreak duration of finite outbreaks
    $T_n$ vs the theoretical prediction in the subcritical and supercritical
    regimes, respectively; (e) and (f) Average outbreak size of finite
    outbreaks $S_n$ vs the theoretical prediction in the subcritical and
  supercritical regimes, respectively.} 
 \label{Tennis}
\end{figure}

\begin{figure}
  \includegraphics[width=\columnwidth]{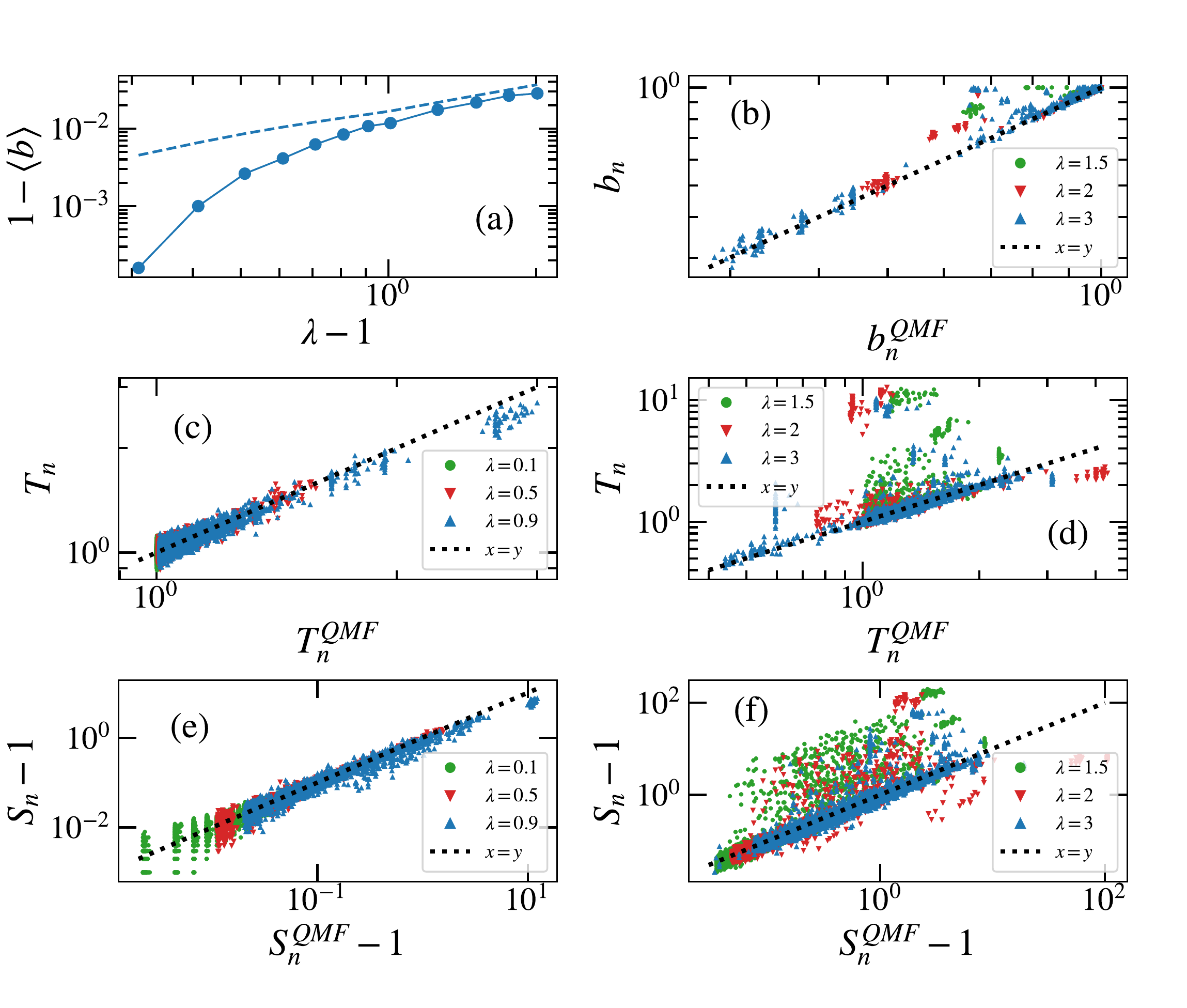} \caption{Numerical
    results for the GR-QC real network. (a) $1-\av{b}$ as a function of
    $\lambda$; (b) Numerical $b_n$ vs the theoretical prediction; (c) and
    (d) Average outbreak duration of finite outbreaks $T_n$ vs the
    theoretical prediction in the subcritical and supercritical regimes,
    respectively; (e) and (f) Average outbreak size of finite outbreaks
    $S_n$ vs the theoretical prediction in the subcritical and supercritical
  regimes, respectively.} 

   \label{GR-QC}
\end{figure}

We have tested the prediction accuracy of our theory on a set of real-world
networks, selected from the list considered in Ref.~\cite{Castellano2017}.
Due to the substantial amount of computer time needed to run simulations in large
networks, we focus on 20 topologies with size between 4000 and 20000 nodes.
As expected the performance of
our theory varies considerably depending on the topology upon which SIS
dynamics occurs. In some cases, the agreement between theory and numerics is
remarkably good.  This is the case of the Tennis network (see
Fig.~\ref{Tennis}), which indeed turns out to be rather uncorrelated and
unclustered.
At the other end of the spectrum is the GR-QC network, Fig.~\ref{GR-QC},
which instead exhibits strong violations of our predictions.

More systematically, we have investigated for all 20 networks considered the
agreement between the predicted value $S_n^{QMF}$ for the average size of
finite outbreaks and the outcome $S_n$ of numerical simulations. We measure
the accuracy of the prediction by evaluating for each network the average
mean relative error of the prediction (MRE), defined as
\begin{equation}
  MRE = \frac{1}{N}\sum_n \left|\frac{S_n}{S_n^{QMF}} -1 \right|,
\label{MRE}
\end{equation}
and correlating it with two typical topological properties present in real
networks but absent in synthetic uncorrelated ones. We consider in
particular the assortativity coefficient $r$, used to measure two-node degree
correlations~\cite{assortative,alexei}, and the average clustering
coefficient $c$, measuring the density of triangles in the network, and thus
its departure from the tree-like assumption~\cite{Watts1998}.

Because of the sampling error, the $MRE$ has a finite expected value
even if the theory is exact (see Appendix~\ref{error}). However, by
increasing the number of sampling averages it is possible to
discriminate whether the measured $MRE$ is truly finite or just
apparently so because of insufficient statistics.

\begin{figure}
  \includegraphics[width=\columnwidth]{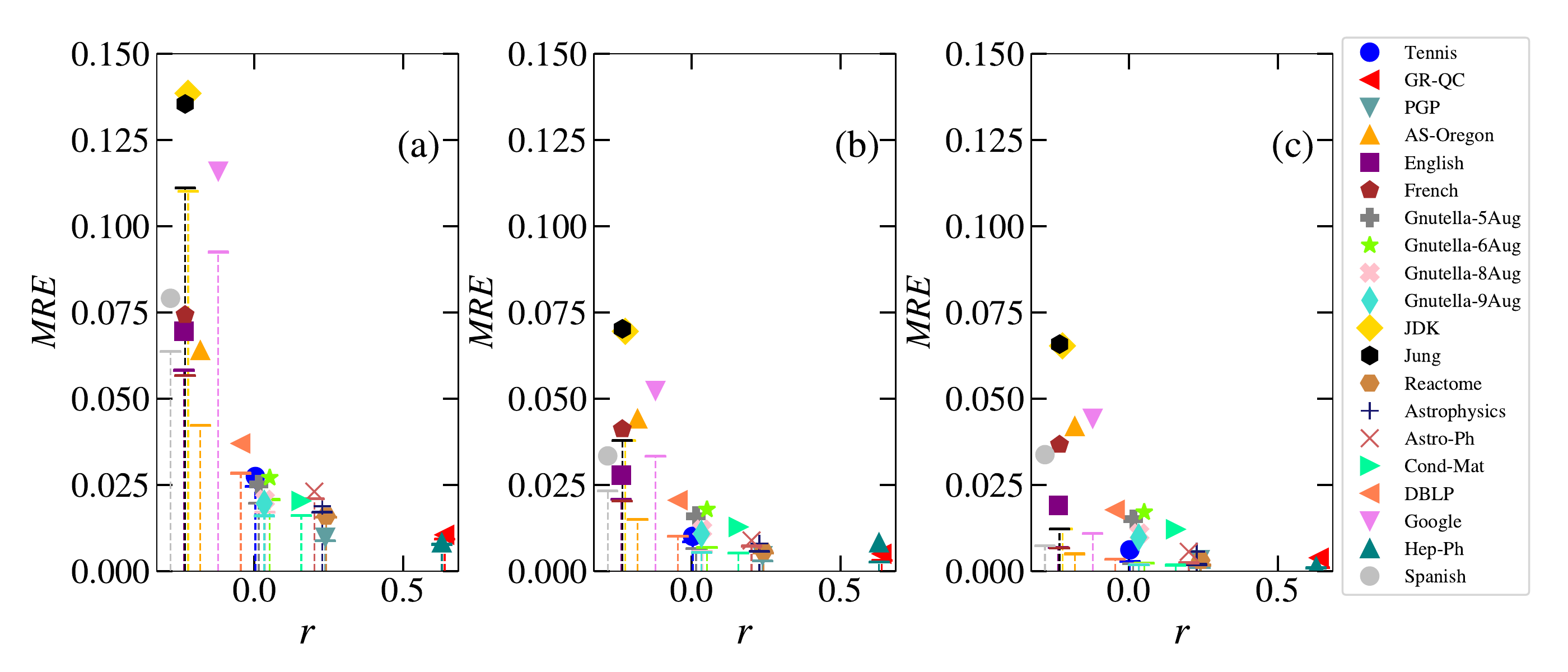} 
  \caption{Mean relative error $MRE$ observed in the set of 20 real
    networks as a function of the assortativity coefficient $r$,
    for $\lambda=0.7$ and increasing number of
    realizations: (a) $N_{r}=10^3$;
    (b) $N_{r}=10^4$; (c) $N_{r}=10^5$. The vertical dashed
    lines denote the expected value, Eq.~(\ref{expMRE}),
    assuming that the theory is exact.}
 \label{MREr}
\end{figure}

\begin{figure}
  \includegraphics[width=\columnwidth]{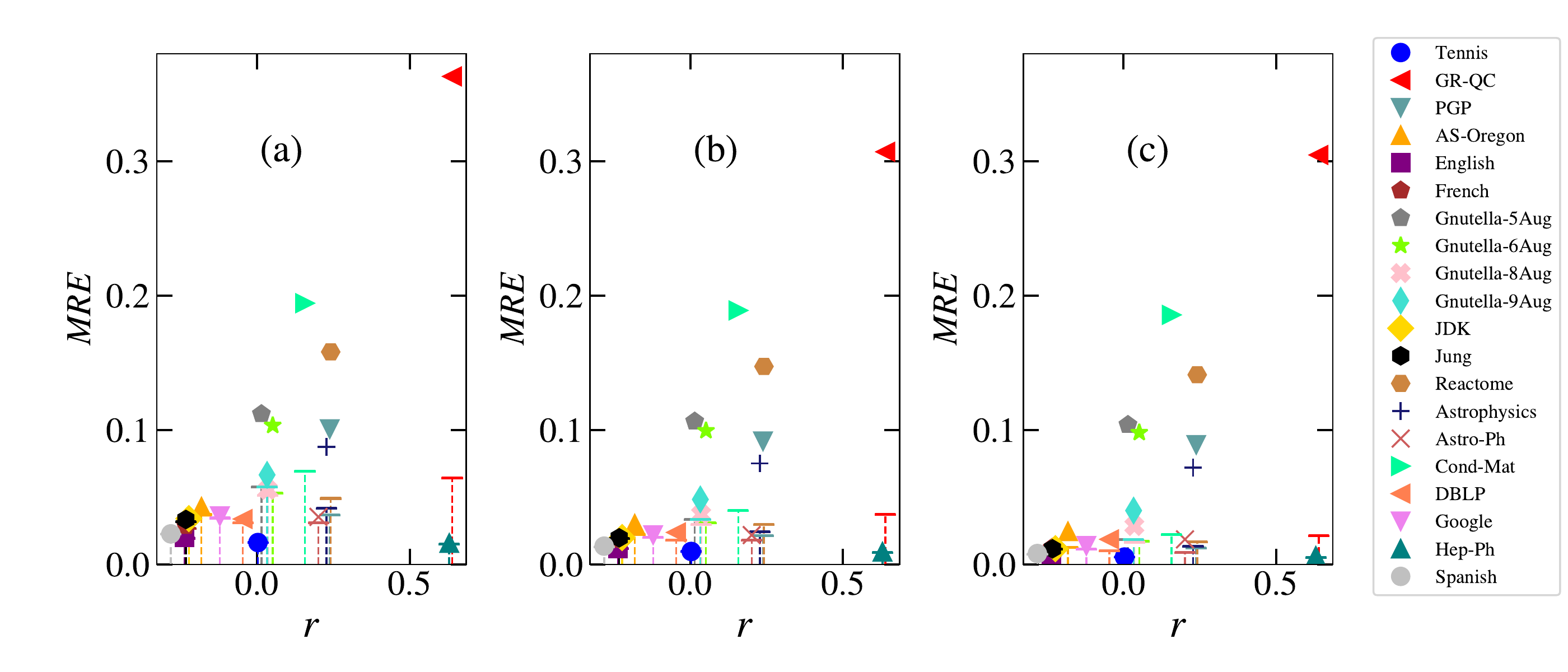} 
  \caption{
    Mean relative error $MRE$ observed in the set of 20 real
    networks as a function of the assortativity coefficient $r$,
    for $\lambda=3$ and increasing number of
    realizations: (a) $N_{r}=10^3$; (b) $N_{r}=3 \times 10^3$; (c) $N_{r}=10^4$.
    The vertical dashed lines denote
    the expected value $MRE_{th}$ assuming that the theory is exact.}
 \label{MREr2}
\end{figure}

In Fig.~\ref{MREr} we report the values of the $MRE$ in real networks as a
function of the assortativity coefficient $r$.
It turns out that $N_{r}=10^4$ is sufficient to reach stationary MRE
values. More importantly, it is apparent that the
average error of the prediction decreases with increasing $r$. This means
that in disassortative networks, corresponding to $r<0$, in which large
degree nodes are preferentially connected to small degree
nodes and viceversa~\cite{assortative}, the average error is larger,
although still limited ($<7\%$).
On the other hand, for assortative networks, with $r>0$, in which
nodes tend to connect with other nodes with similar degree, the average
error is practically vanishing.
For supercritical values of $\lambda$ (see Fig.~\ref{MREr2}) , instead,
an opposite behavior emerges. The errors tend to vanish for
disassortative networks while they remain quite large for many networks
with positive $r$.

In order to explore more deeply the effects of degree correlations, we
consider random networks with given degree distribution and degree
correlations, generated according to the Weber-Porto
(WP)~\cite{PhysRevE.76.046111} prescription.
\begin{figure}
  \includegraphics[width=\columnwidth]{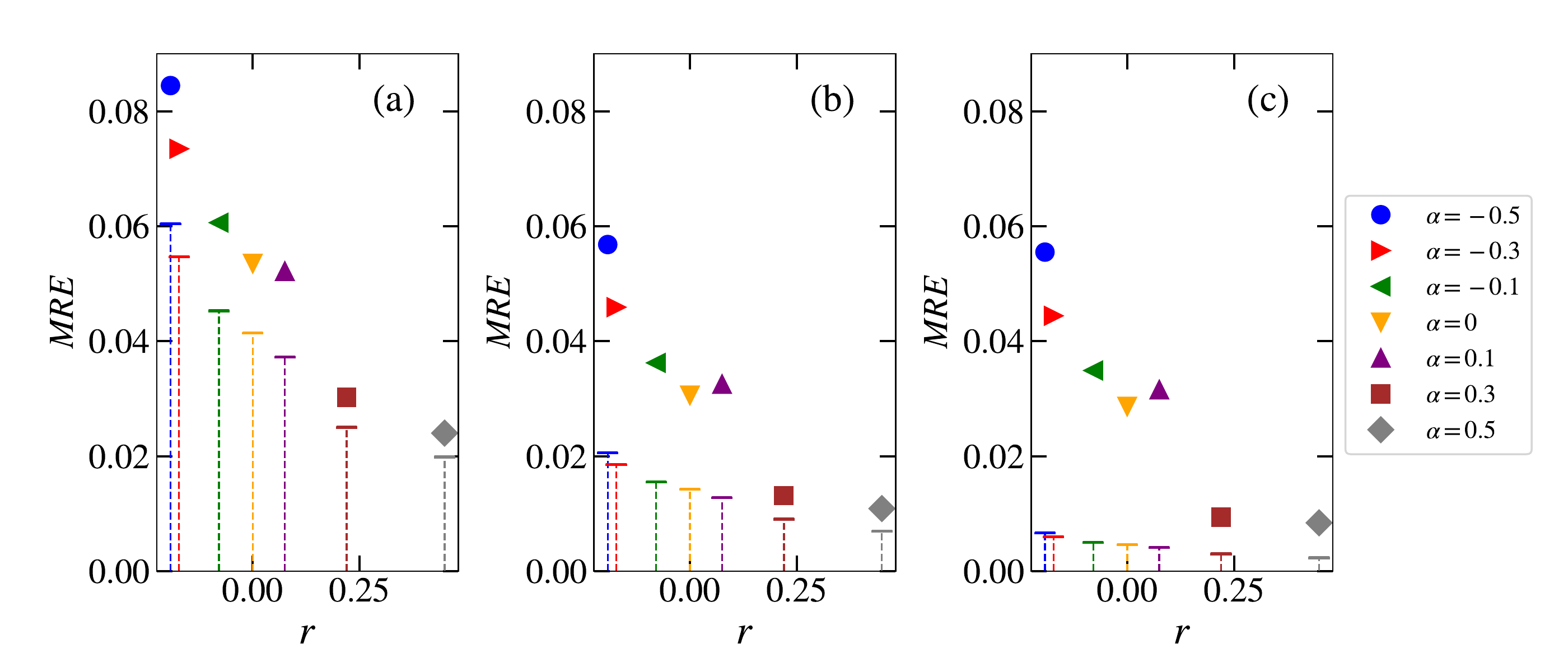}
  \caption{Mean relative error $MRE$ observed in Weber-Porto networks
    as a function of the assortativity coefficient $r$, for
    $\lambda=0.7$ and increasing number of realizations: (a)
    $N_{r}=10^3$; (b) $N_{r}=10^4$; (c) $N_{r}=10^5$. The vertical
    dashed lines denote the expected value $MRE_{th}$ assuming that
    the theory is exact.  Networks used have size $N=10^4$ and degree
    exponent $\gamma=2.25$.}
 \label{MRE_WP}
\end{figure}
In this model, degree correlations are defined in terms of the
average degree of the nearest neighbors of nodes of degree $k$,
$\bar{k}_{nn}(k)$, which is an increasing function for assortative
correlations and a decreasing one for disassortative ones~\cite{alexei02}.
Choosing a form $\bar{k}_{nn}(k) \sim k^{\alpha}$, where $\alpha<0$
($\alpha>0$) corresponds to disassortative (assortative) networks, we find
for the subcritical case the results shown in Fig.~\ref{MRE_WP}.
As we can see, WP networks in the subcritical regime behave in a manner
analogous to real networks, with a MRE decreasing with increasing $r$.
\begin{figure}
  \includegraphics[width=\columnwidth]{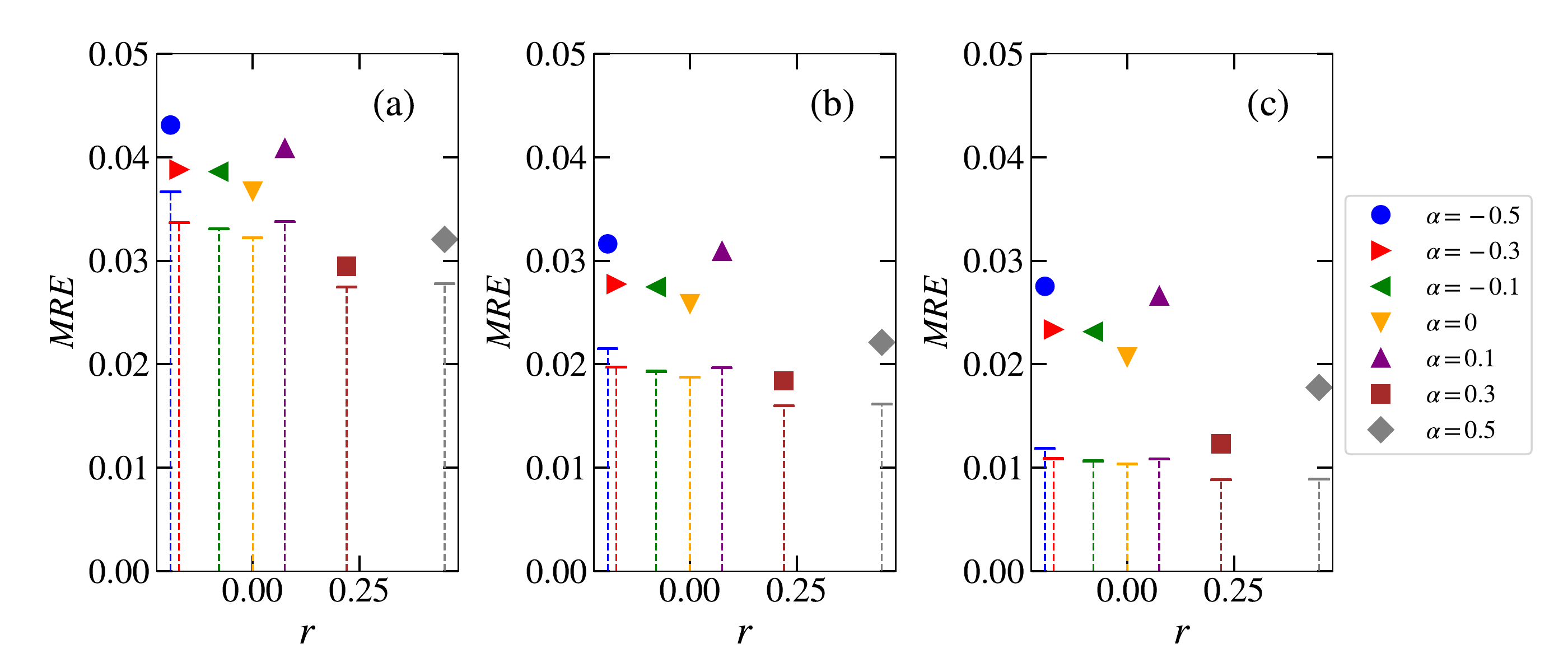}
  \caption{Mean relative error $MRE$ observed in Weber-Porto networks
    as a function of the assortativity coefficient $r$, for
    $\lambda=3$ and increasing number of realizations: (a)
    $N_{r}=10^3$; (b) $N_{r}=3 \times 10^3$; (c) $N_{r}=10^4$. The vertical
    dashed lines denote the expected value $MRE_{th}$ assuming that
    the theory is exact.  Networks used have size $N=10^4$ and degree
    exponent $\gamma=2.25$.}
 \label{MRE_WP2}
\end{figure}
In the supercritical regime, on the other hand (see Fig.~\ref{MRE_WP2}),
the relative errors are practically independent of correlations.
This observation is not in agreement with the phenomenology observed
in real networks.

\begin{figure}
  \includegraphics[width=\columnwidth]{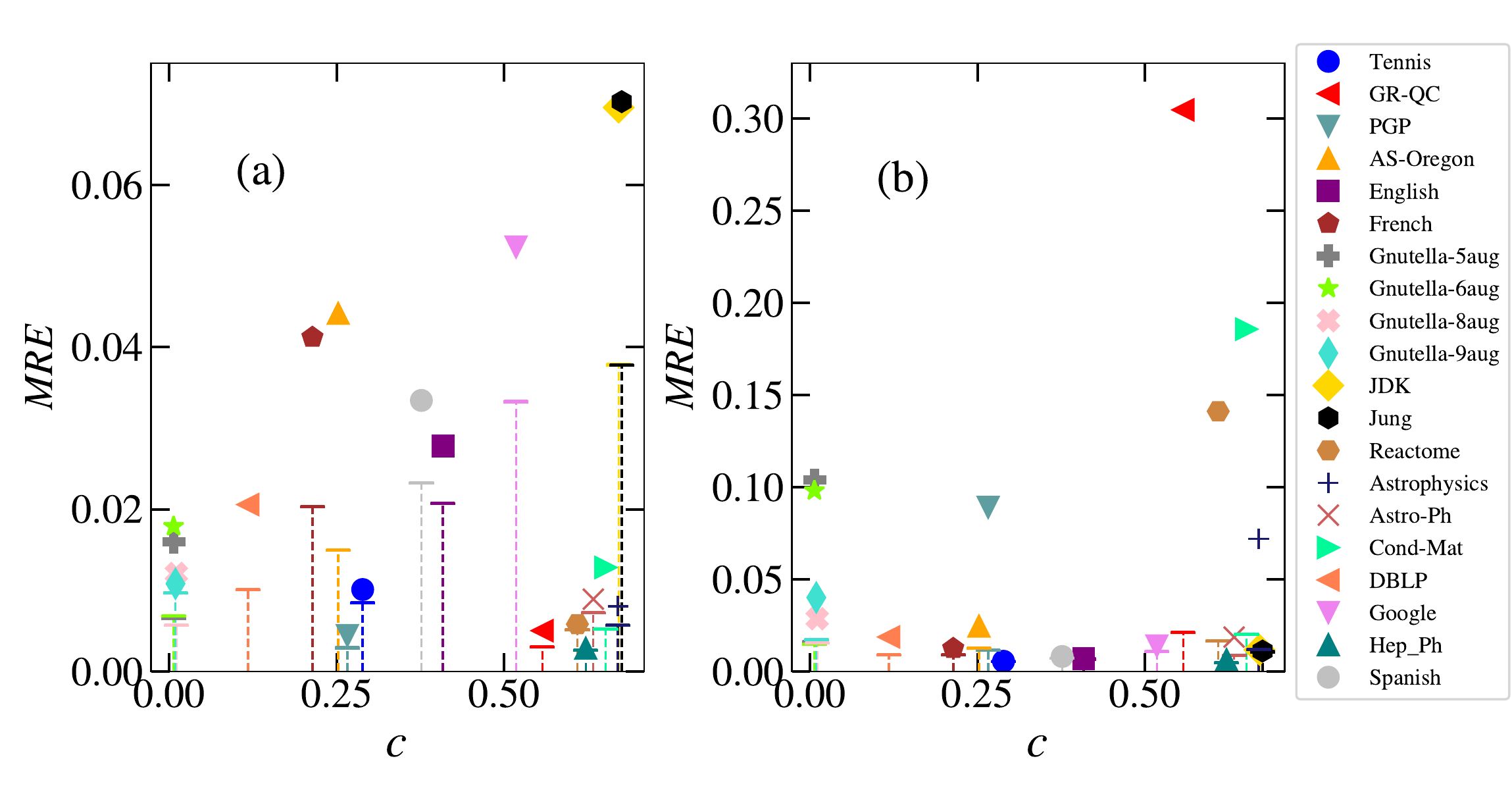}
  \caption{
    Mean relative error $MRE$ observed in the set of 20 real
    networks as a function of the clustering coefficient $c$,
    for $N_r=10^4$ and $\lambda=0.7$ (left), $\lambda=3$ (right).
    The vertical dashed lines denote the expected value $MRE_{th}$
    assuming that the theory is exact}
 \label{MREc}
\end{figure}

Finally, in Fig.~\ref{MREc} we report the values of MRE as a
function of the average clustering coefficient $c$ in our set of real
networks. In this case, no clear dependence on $c$ can be identified.
These results suggest that the presence of strong disassortative
degree correlations reduces the performance of our theory while other
topological features (different from local clustering) are responsible for
the other discrepancies between our theory and numerical simulations in
real networks.

It is important to remark  that the values of MRE indicate nevertheless a
rather impressive overall accuracy of the theory.
The relative error is smaller than $20\%$ for all networks but one
(for which is a still acceptable $31\%$),
indicating that even in real-world networks,
with all their intricacies, our prediction is a good baseline for
estimating the spreading influence of individual nodes.

\section{Conclusions}

In this paper we have developed a theory that allows to calculate, for each
single node $n$ in a network, the probability that a continuous-time
SIS outbreak started in
$n$ remains finite, its average duration and size.  The theoretical
treatment is based on a QMF-type approach to SIS dynamics and hence it
shares strengths and weaknesses with the latter.  For strongly heterogeneous
uncorrelated networks with degree exponent $\gamma<5/2$ the theory works
very well and is asymptotically exact for large networks.  For $\gamma>5/2$
instead it constitutes only an approximation, whose accuracy is good only
for not too large systems.  The nontrivial features of the SIS epidemic
transition, even on uncorrelated networks, result in a dependence of the
spreading influence on topological features different from the simple degree
centrality: Depending on the value of the degree exponent $\gamma$
the K-core index or the distance from the largest hub play a
relevant role.  When applied to SIS dynamics on real-world networks our
theory turns out to be rather surprisingly accurate, with deviations from it
more related to degree-correlations than to clustering.

When assessing the validity of the present approach it is to be remarked
that we are able to make reasonably precise predictions for the values of
the observables, with no fitting parameter to be adjusted.  This is to be
contrasted with many other approaches for the identification of influential
spreaders for SIR dynamics, where the (much more limited) goal is the
assessment of whether a node is a better spreader than another, with no
prediction for the actual outbreak size or duration.

Despite this success, there is clearly still room for improvement.  A first
avenue of research should attempt to improve the theory in order to better
predict the behavior for uncorrelated networks with $\gamma>5/2$.  Theories
slightly improving on QMF, such as pair-quenched mean-field~\cite{Mata2013},
have been proposed, but in order to fully capture the complexity of the SIS
epidemic transition for $\gamma>5/2$ one has to consider a long-range
percolation process~\cite{Castellano2020}, which appears not easily
applicable to the calculation of the spreading influence.

The other natural continuation of the present research is in the direction
of better understanding spreading influence for real-world networks.
It is likely that progress in this direction will require the consideration
of other suitable synthetic network models, allowing to understand one at a time
the effect of the various topological properties.

\begin{acknowledgments}
  We acknowledge financial support from the Spanish MICINN, under Project
  No. PID2019-106290GB-C21.
\end{acknowledgments}

%

\appendix

\section{Fixed points of $\mathbf{b_n}$}
\label{cap_fixed_points}

In order to study the possible steady state values of the probability
$b_n$, we perform a linear stability analysis of the differential equation
Eq.~\eqref{eq_c}. The constant value $b_n = 1$ is always a solution of
Eq.~\eqref{eq_c}. Let us assume a small variation from this value, defined
as $c_n(t) = 1 - \epsilon_n(t)$. Introducing this expression into
Eq.~\eqref{eq_c} and keeping only the lowest order terms, we find
\begin{equation}
  \frac{d \epsilon_n(t)}{dt} \simeq -\mu \epsilon_n(t) + \beta \sum_m a_{nm}
  \epsilon_m(t)
  = \sum_m L_{nm} \epsilon_m(t)  \nonumber
\end{equation}
defining the Jacobian matrix
\begin{equation}
  L_{nm} = -\mu \delta_{nm}  + \beta a_{nm}.
\end{equation}
The solution $b_n = 1$ is stable if the largest eigenvalue of the
Jacobian matrix is negative. Given its structure, we can readily see that
the eigenvectors of the adjacency matrix are also eigenvectors of $L_{nm}$,
with an associated eigenvalue $\Lambda_L = -\mu + \beta \Lambda_A$. The
largest eigenvalue of $L_{nm}$ is thus $-\mu + \beta \Lambda_M$, and the
condition for the stability of the solution $b_n = 1$ is 
\begin{equation}
  -\mu + \beta \Lambda_M < 0 \quad \Rightarrow \quad \frac{\beta}{\mu} \Lambda_M \equiv
  \lambda < 1.
\end{equation}
For $\lambda > 1$ the solution $b_n = 1$ becomes unstable, and the steady
state is given by the new stable solution $b_n < 1$ obtained from the
recursive relation Eq.~\eqref{eq_b_qmf}.

\section{Unphysical nature of the linear approximation for the average
outbreak time}
\label{unphysical}

The linear approximation for the quantity $f_n(t) = b_n - c_n(t)$, obtained
by neglecting the quadratic terms in Eq.~\eqref{eq_df}, takes the form in
rescaled time
\begin{equation}
  \frac{d f^L_n(t)}{dt} = -  \frac{f^L_n(t)}{b_n} + \alpha b_n \sum_m a_{nm}
  f^L_m(t),
  \label{eq:appen_unphys}
\end{equation}
where $\alpha = \beta / \mu$.  By its very definition, the probability
$c_n(t)$ must be an increasing function in the interval $[0,\infty]$.
Consequently, the function $f_n(t)$ must be a decreasing function in the
same interval, that is, $\frac{f_n(t)}{dt} < 0$ for $t \geq 0$, and, 
with the initial condition $f_n(0) = b_n$, we must have $f_n(t) \leq b_n$
for $t \geq 0$.

From Eq.~\eqref{eq:appen_unphys}, we can compute the slope of the function
$f^L_n(t)$ at time $t=0$ given by
\begin{eqnarray}
  \frac{d f^L_n(0)}{dt} 
  &=& - \frac{f^L_n(0)}{b_n} + \alpha b_n \sum_m a_{nm} f^L_m(0) \nonumber \\
  &=& - 1 + \alpha b_n \sum_m a_{nm} b_m \nonumber \\
  &=& b_n(1 + \alpha k_n)- 2, \label{eq:append_unphys_2}
\end{eqnarray}
where we have used the steady state condition Eq.~\eqref{eq_b_qmf}.  From
Eq.~\eqref{eq:append_unphys_2} we can see that, for nodes that fulfill the
condition 
\begin{equation}
  b_n > \frac{2}{1 + \alpha k_n}
\end{equation}
$f^L_n(t)$ is an initially increasing function with positive derivative,
taking at short times values larger than $b_n$. This situation is completely
unphysical, since it would imply that the probability $c_n(t)$ is negative.
We therefore conclude that the linearized equation for $f_n(t)$ is
unphysical, and cannot be used to compute the average outbreak size. This is
the situation, in particular, of nodes with large degree.

\begin{figure}[t]
  \centering
  \includegraphics[width=\columnwidth]{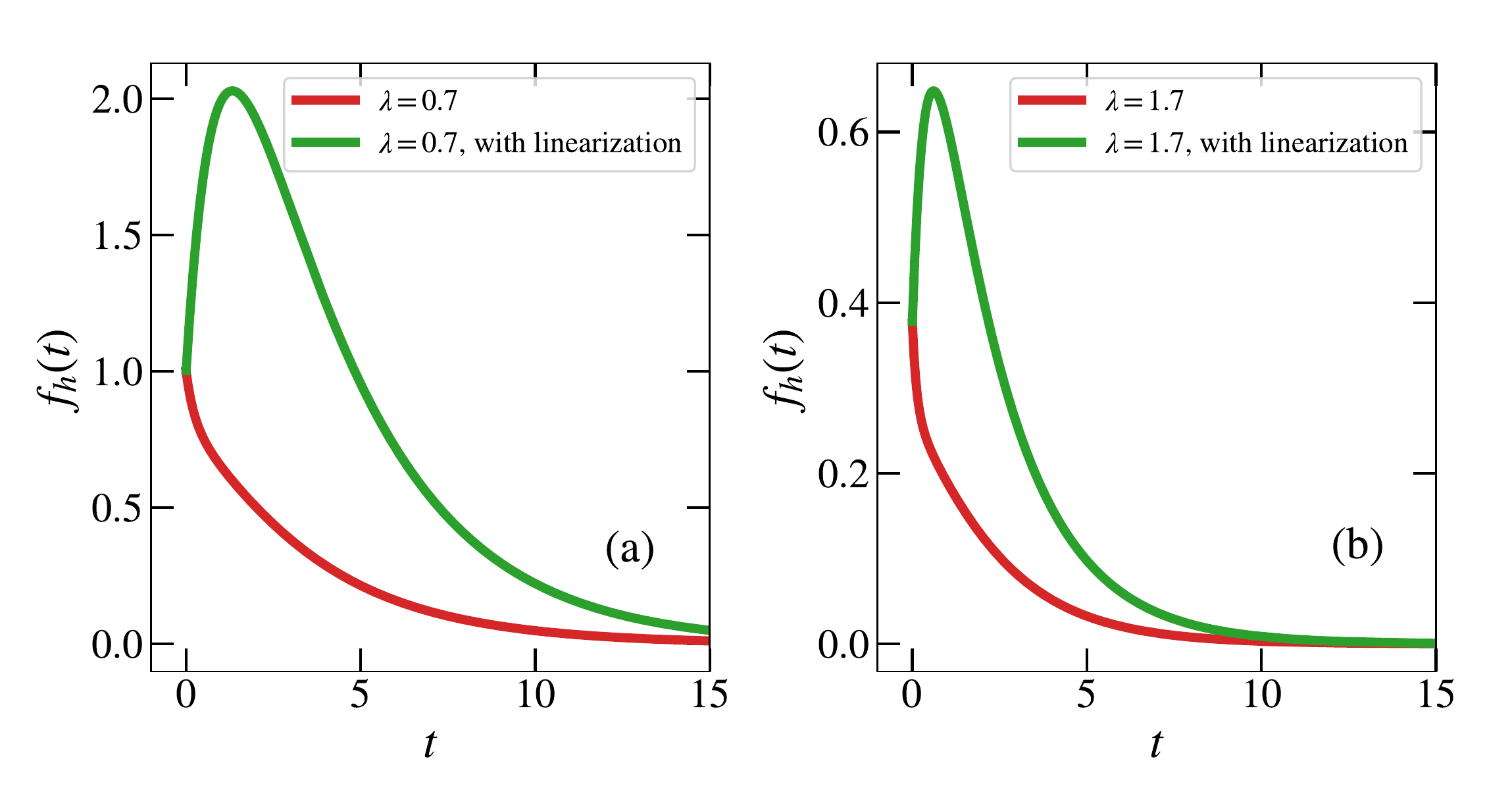}

  \caption{Full function $f_n(t)$ and its analogous from the linear
    approximation $f_n^L(t)$ numerically computed for the hub (node with
    maximum degree) in a network of size $N=10^4$ and degree exponent
    $\gamma = 2.25$. (a) Subcritical regime, with  $f_h(0) = 1$. (b)
  Supercritical regime, with $f_h(0) = b_n < 1$.}
  \label{non-decreasing}
\end{figure}

In Fig.~\ref{non-decreasing} we report the results of the numerical
integration of the linealized equation, compared with the numerical
integration of the full nonlinear equation Eq.~\eqref{eq_df}. As we can see,
for large values of the degree, the linearized function $f^L_n(t)$ shows a
characteristic maximum in the vicinity of $t=0$, which is absent in the full
non-linear solution.

\section{The distinction between finite and infinite outbreaks in simulations}
\label{choice}

\begin{figure}[t]
\centering \includegraphics[width=\columnwidth]{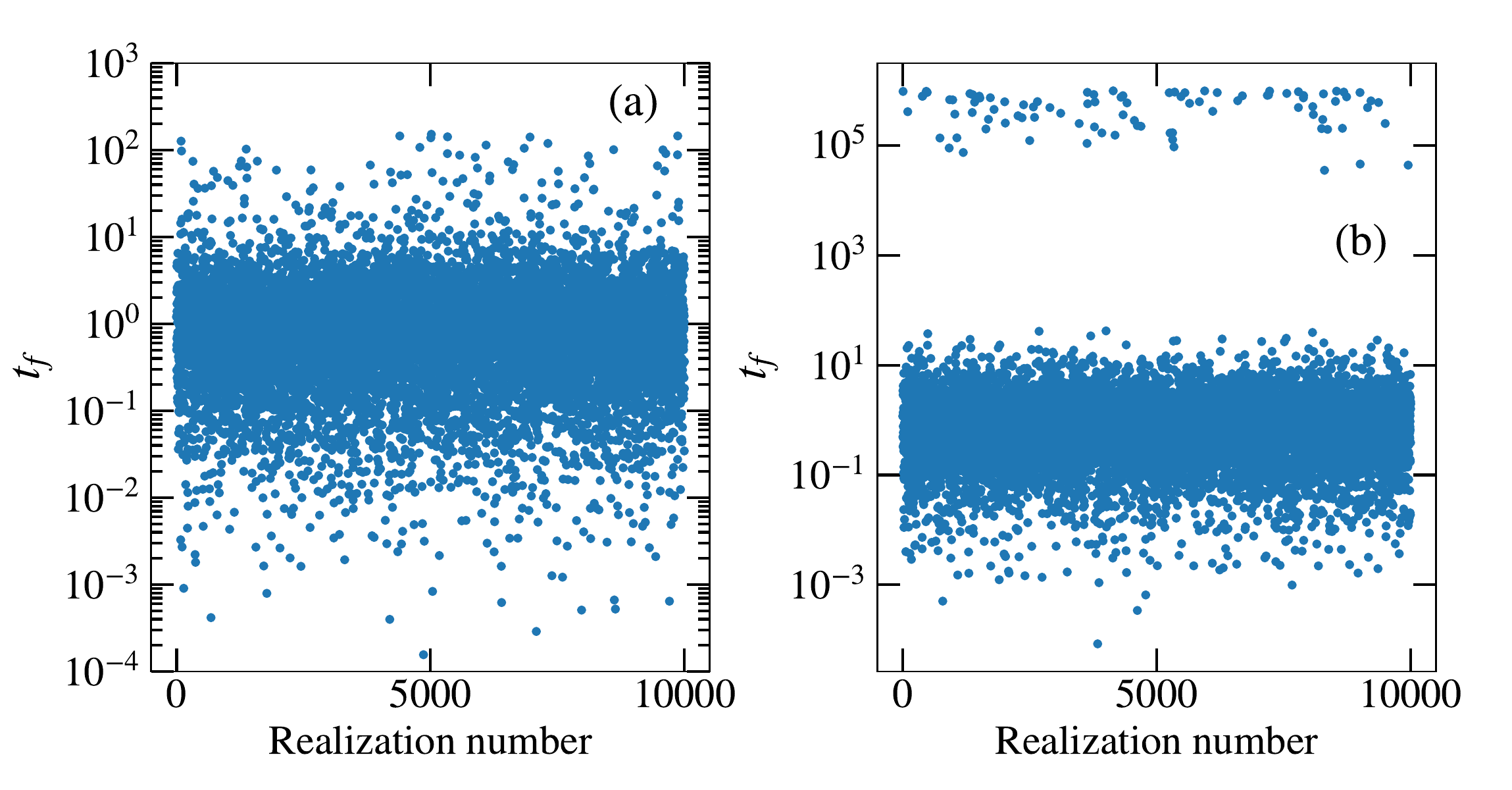}
\caption{Total outbreak duration $T_f$ in many realizations of the outbreak
  process in network with $\gamma=2.25$, $N=10^4$, $\lambda=1.1$ (left panel)
  and $\lambda=1.3$ (right panel).}
\label{Tfin}
\end{figure}

Since we are dealing with simulations in finite size systems we must decide
a criterion to distinguish (above the epidemic threshold) between truly
finite outbreaks (upon which we are performing averages) and outbreaks
which would give rise to the stationary state in an infinite system, but that,
in the finite networks we consider, end only because of fluctuations. In order
to determine such a criterion, we look at the total duration $t_f$ of
outbreaks for several realizations of the epidemic process, see
Fig.~\ref{Tfin}.   While close to the transition the distribution of
$t_{f}$ is singly-peaked, if $\lambda$ is increased two well separated components
appear, one corresponding to small values of $t_f$, which represent truly
finite outbreaks, and the other for extremely large values of $t_f$, that we
can associate to putatively infinite outbreaks that are stopped
only due to finite
size effects. In the presence of this gap, it is possible to
define a maximum duration $T$, located between the two
clusters of points, to distinguish the two types of outbreaks.
A duration $t_f < T$ implies a finite outbreak, while $t_f \geq T$
means that the outbreak must be considered infinite. In view of
Fig.~\ref{Tfin}, we set $T = 100$ throughout all simulations reported
in the paper.  For larger values of $N$, clearly it would be possible to
operate the distinction even for values of $\lambda$ closer to 1, at the
price of increasing $T$ and hence the simulation time.

\section{Expected value of the $MRE$}
\label{error}

When comparing simulations with theory, even in the case the latter is exact,
i.e., the expected value of the outbreak size is $S_n^{QMF}$,
sampling error implies that the $MRE$ (Eq.~\eqref{MRE}) has a finite
value depending on the number of realizations considered.
Indeed, for $N_{r}$ realizations, from the central limit theorem
\be
S_n = S_n^{QMF} + \eta_n
\ee
where $\eta_n$ is a Gaussian $P(\eta_n)$ of zero mean and variance
$\sigma_n^2/N_{r}$, with $\sigma_n^2$ the variance of outbreak size
distribution for given $n$. Then the expected $MRE$ value is
\begin{eqnarray}
MRE_{th} & = &\frac{1}{N} \sum_n \frac{1}{S_n^{QMF}} \int_{-\infty}^{+\infty} d\eta_n P(\eta_n) |\eta_n| \\ \nonumber
&=& \frac{2}{N} \sum_n \frac{1}{S_n^{QMF}} \int_{0}^{\infty} d\eta_n P(\eta_n) \eta_n \\ \nonumber \label{expMRE}
&=& \frac{2}{\sqrt{2 \pi N_{r}}} \frac{1}{N} \sum_n \frac{\sigma_n}{S_n^{QMF}}
\end{eqnarray}
Therefore, by increasing $N_{r}$ it is possible to discriminate whether
the measured $MRE$ is truly finite or just because of insufficient statistics.

\end{document}